# Adaptation dynamics in densely clustered chemoreceptors

William Pontius[1,2]

Michael W. Sneddon[2,3,4]

Thierry Emonet[1,2]

[1]Department of Physics, Yale University, New Haven, CT, USA

[2]Department of Molecular, Cellular, and Developmental Biology, Yale University, New Haven, CT, USA

[3]Interdepartmental Program in Computational Biology and Bioinformatics, Yale University, New Haven, CT, USA

[4]Current address: Physical Biosciences Division, Lawrence Berkeley National Lab, Berkeley, CA, USA

In many sensory systems, transmembrane receptors are spatially organized in large clusters. Such arrangement may facilitate signal amplification and the integration of multiple stimuli. However, this organization likely also affects the kinetics of signaling since the cytoplasmic enzymes that modulate the activity of the receptors must localize to the cluster prior to receptor modification. Here we examine how these spatial considerations shape signaling dynamics at rest and in response to stimuli. As a model system, we use the chemotaxis pathway of *Escherichia coli*, a canonical system for the study of how organisms sense, respond, and adapt to environmental stimuli. In bacterial chemotaxis, adaptation is mediated by two enzymes that localize to the clustered receptors and modulate their activity through methylation-demethylation. Using a novel stochastic simulation, we show that distributive receptor methylation is necessary for successful adaptation to stimulus and also leads to large fluctuations in receptor activity in the steady state. These fluctuations arise from noise in the number of localized enzymes combined with saturated modification kinetics between the localized enzymes and the receptor substrate. An analytical model explains how saturated enzyme kinetics and large fluctuations can coexist with an adapted state robust to variation in the expression levels of the pathway constituents, a key requirement to ensure the functionality of individual cells within a population. This contrasts with the well-mixed covalent modification system studied by Goldbeter and Koshland in which mean activity becomes ultrasensitive to protein abundances when the enzymes operate at saturation. Large fluctuations in receptor activity have been quantified experimentally and may benefit the cell by enhancing its ability to explore empty environments and track shallow nutrient gradients. Here we clarify the mechanistic relationship of these large fluctuations to well-studied aspects of the chemotaxis system, precise adaptation and functional robustness.

# Introduction

High-resolution microscopy has revealed the exquisite spatial organization of signaling pathways and their molecular constituents. Understanding the computations performed by biological networks therefore requires taking the spatiotemporal organization of the reactants into account [1]. One feature common to many signal transduction pathways is the clustering of receptors in the cell membrane. This arrangement has been observed for diverse receptor types [2] such as bacterial chemoreceptors [3-6], epidermal growth factor receptors [7], and T cell antigen receptors [8]. Receptor clustering provides a mechanism for controlling the sensitivity [9,10] and accuracy [11,12] of a signaling pathway. Moreover, by controlling which types of receptors participate in clusters a cell can achieve spatiotemporal control over the specificity of the signaling complexes.

While clustering receptors can tune the sensitivity and specificity of a signaling pathway, organizing receptors into clusters also imposes novel constraints on the kinetics of the pathway. Temporal modulations of the activity of signaling complexes, such as adaptation, are typically achieved via posttranslational modification of the cytoplasmic tail of the receptors by various enzymes. The localization of the receptor substrate into clusters implies that trafficking of enzymes between the cytoplasm and the cluster and between receptors within a cluster is likely to be an important determinant of the dynamics of such modulations. Recent theoretical studies of the effect of the localization of enzymes and substrates on signaling kinetics have shown that spatiotemporal correlations between reactants can significantly affect the signaling properties of these pathways [13-15].

One well-characterized system in which the spatial organization of receptors plays a significant role is the chemotaxis system of the bacterium *Escherichia coli* [16-18]. *E. coli* moves by performing a random walk alternating relatively straight runs with sudden changes of direction called tumbles. The probability to tumble is modulated by a two-component system in which transmembrane receptors regulate the activity of a histidine kinase CheA, which in turn phosphorylates the response regulator CheY. Phosphorylated CheY rapidly diffuses through the cell and binds the flagellar motors to induce tumbling. The tumbling rate decreases in response to chemical attractants and increases in response to repellants, allowing the bacterium to navigate its environment.

Chemoreceptor clustering affects both signal amplification and adaptation to persistent stimuli, which together enable bacteria to remain sensitive to over five orders of magnitude of ligand concentration [19]. Signal amplification arises from allosteric interactions between clustered receptors [9,20-23] whereas adaptation is mediated by the activity of two enzymes: CheR methylates inactive receptors, thereby reactivating them, while CheB demethylates active receptors, deactivating them. This arrangement implements an integral feedback mechanism [24], enabling kinase activity and therefore cell behavior to return to approximately the same stationary point following response to stimulus [25,26]. The localization of enzymes to the cluster is facilitated by a high-affinity tether site present on most receptors. This tether, together with the dense organization of the cluster, enables localized enzymes to modify multiple receptors within a range known as an assistance neighborhood [27]. Modeling efforts have shown that assistance neighborhoods are required for precise adaptation when receptors are strongly coupled [28].

Precise adaptation, however, is not by itself sufficient for successful chemotaxis. The dynamics of the adaptation process, including the rate of receptor modification and the level of spontaneous fluctuation in receptor activity, are also critical determinants of chemotactic performance [29-35]. Recent measurements of the dynamic localization of chemotaxis proteins have shown that the time scale of CheR and CheB localization to the receptor cluster is comparable to the time scale of adaptation [36] and therefore expected to affect the dynamics significantly. Moreover, dense clustering may enable localized enzymes to perform a random walk over the receptor lattice without returning to the cytoplasmic bulk, a proposed process termed brachiation [37] that may lead to more efficient receptor modification.

Here we analyze how the spatiotemporal localization of the adaptation enzymes to the receptor cluster affects the dynamics of the adaptation process. First we build a stochastic simulation of the chemotaxis system taking into account the organization of the receptors into large clusters [4,6], the slow exchange of enzymes between the cytoplasm and the clusters [36], enzyme brachiation [37], and assistance neighborhoods [27,28,38]. This model quantitatively recapitulates experimental observations of the magnitude of the spontaneous fluctuations in single cells [39-42] and the kinetics of adaptation averaged over multiple cells [43]. Notably, while localized enzymes in this model operate at saturation, the output of the system nonetheless remains robust to cell-to-cell variation in enzyme expression levels [44], in contrast to the covalent modification system studied by Goldbeter and Koshland [12]. We therefore resolve the question of how large spontaneous fluctuations might coexist with a robust mean output in the system [30]. We interpret these results in the second part of the paper, using a mean-field analytical model to examine the molecular mechanisms underlying these features and their relation to receptor clustering.

# Results

**Numerical model of adaptation dynamics in a chemoreceptor cluster**

We used the rule-based simulation tool NFsim [45] to create a stochastic model of the bacterial chemotaxis system that accounts for the organization of chemoreceptors into a large, dense, hexagonal lattice [4]. Like the Gillespie algorithm, NFsim computes exact stochastic trajectories, but avoids the full enumeration of the reaction network, which can undergo combinatorial explosion, by using rules to generate reaction events [45]. In the simulation, each chemoreceptor dimer is represented by an object with one tether site, one modification site, and a methylation level ranging from 0 to 8. We model a single contiguous lattice consisting typically of 7200 dimers, although we consider different sizes as well. The structure of the lattice is fully specified by enumerating for each dimer its six nearest neighboring dimers. Receptor cooperativity is modeled using Monod-Wyman-Changeux (MWC) complexes consisting of six receptor dimers (Fig. 1A). The activity $a$ of each signaling complex depends on the methylation and ligand-binding state of the dimers in the complex and is calculated from Eq. (13) (Methods) as previously described [23,28]. The implementation of this model in NFsim is discussed in the Supporting Text S1.

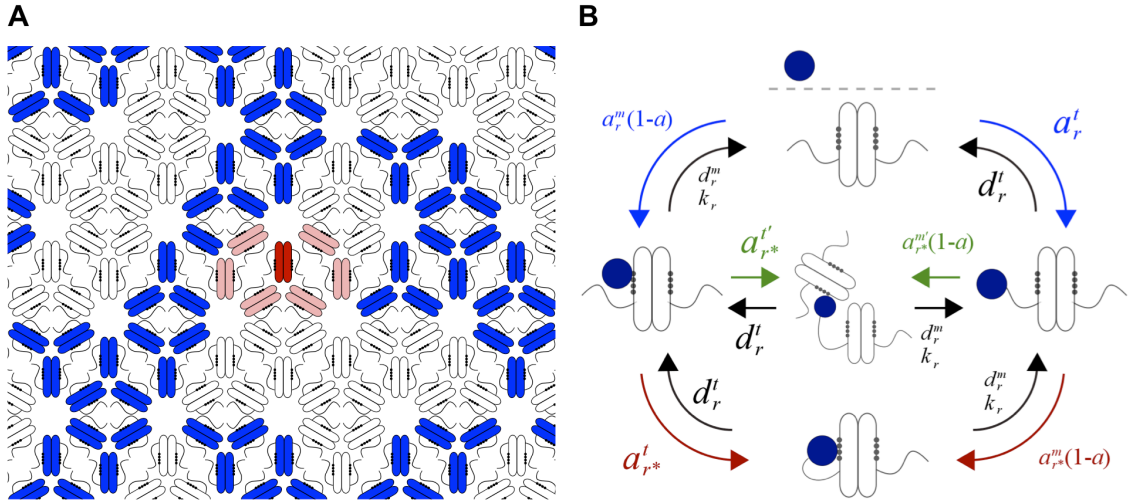

**Fig. 1**: Adaptation reactions on the chemoreceptor lattice. (A) Bacterial chemoreceptors assemble into trimers of dimers that organize to form a dense hexagonal lattice. Most chemoreceptors have tether and modification sites. In the model, the assistance neighborhood for a given receptor (red) consists of all the receptors accessible by its tether, here taken to be the six nearest dimers (light red) in addition to itself. Groups of six receptor dimers switch cooperatively between active (blue) and inactive (white) states according to a MWC model. (B) Modeled reactions between CheR and the chemoreceptors with corresponding rates. Binding rates to the modification site depend on the receptor activity $a$. CheR in the cytoplasmic bulk may bind either the tether or modification site of a receptor (blue arrows, rates $a_r^t$ and $a_r^m(1-a)$ respectively). Once bound to the tether or modification site it may respectively bind the modification site or tether of itself (red arrows, rates $a_{r*}^m(1-a)$ and $a_{r*}^t$ respectively) or any other receptor within its assistance neighborhood (green arrows, rate $a_{r*}^{m'}(1-a)$ to bind the neighboring modification site and rate $a_{r*}^{t'}$ to bind the neighboring tether). Black arrows denote unbinding and catalytic steps (catalytic rate $k_r$; tether unbinding rate $d_r^t$; modification site unbinding rate $d_r^m$). CheB-P participates in analogous reactions. In the rates, superscripts $m$ and $t$ denote binding to the modification site and tether site, respectively. The subscripts $r$ and $b$ denote CheR and CheB reactions, respectively.

Receptor modification occurs through the enzymes CheR and CheB, which are each modeled as having two binding sites, one specific to the receptor tether and one specific to the modification site. In the model, CheR and CheB dynamically bind and unbind both of these sites. CheR participates in the reactions illustrated in Fig. 1B. The possible states of the enzyme are: free and dispersed in the cytoplasmic bulk, or bound to one or both of the tether and modification sites. Enzymes in the bulk localize to the cluster by binding either the tether site or the modification site directly. The time scales of these binding reactions (Fig. 1B, blue arrows) are the slowest in the present model: ~15s for localization through tether binding, as measured [36], and longer for modification site binding, reflecting the lower affinity of enzymes for the modification site. Once bound to the tether or modification site, an enzyme may bind the modification site or tether, respectively, of the receptor to which it is already bound (Fig. 1B, red arrows) or any of its six nearest neighbors (green arrows). Therefore the assistance neighborhood consists of seven dimers, consistent with measurements [27]. Assistance neighborhoods are unique for each receptor dimer and therefore overlap. Accordingly, in the simulation individual receptor dimers participate in multiple assistance neighborhoods. Since these reactions are confined to small volumes (given by the ~5 nm tether radius [46]), they proceed at high rates (1-10 ms time scales; see Text S1). The activity-dependent binding rate of CheR to the modification site is proportional to $1 - a$, while the rates of all other CheR reactions are taken to be independent of activity. Phosphorylated CheB (CheB-P) participates in completely analogous reactions except that the rate of binding the modification site is proportional to $a$. CheB phosphorylation proceeds at a rate proportional to the activity of the receptor cluster (Text S1). For simplicity we assume that only CheB-P can localize to the receptor cluster since its affinity for the tether is much higher than that of CheB [47].

Our study is the first to incorporate enzyme brachiation [37], assistance neighborhoods [28,38], cooperative amplification of the input signal [9,22,23], activity-dependent adaptation kinetics [25], and a large contiguous receptor cluster into a single model. This model specifically extends two earlier models. The first of these models considered enzyme brachiation on a large receptor cluster [37], but did not include activity-dependent kinetics, receptor cooperativity, or any modification of the receptors. The second of these models included activity-dependent kinetics, cooperativity, and assistance neighborhoods [28,38] but excluded enzyme brachiation and limited the system size to a single MWC complex consisting of 19 dimers. Here we take advantage of the flexibility and efficiency of NFsim to examine how all of these processes together determine the dynamics of adaptation.

Calibration of the model parameters is discussed in the Supporting Text S1. Supporting Tables S1 and S2 present the full set of simulation parameters. We note that our model includes only Tar receptors. This choice enabled us to compare our model directly to measurements of the adaptation kinetics [43] performed on cells lacking receptors other than Tar. These measurements were obtained by exposing cells to time-dependent exponential ramps of methyl-aspartate, a protocol that we modeled *in silico* (Fig. 2A and Fig. S2) to verify the calibration of the kinetics of our model. In the remainder of the paper we denote this calibrated model as the reference model **M1**.

**Distributive methylation leads to precise adaptation**

Together with the dense organization of the receptor lattice, the presence of the tether site on each receptor gives rise to assistance neighborhoods [27] and possibly enzyme brachiation [37]. During the brachiation process, enzymes successively bind and unbind the tethers and modification sites on different, neighboring receptors, enabling them to perform a random walk over the lattice without returning to the bulk. Both assistance neighborhoods and enzyme brachiation should increase the distributivity of the methylation process, meaning that sequential (de)methylation events catalyzed by a single enzyme will tend to take place on different receptors. In a distributive scheme, therefore, an enzyme will tend to modify multiple receptors during its residence time on the cluster. Moreover, it will tend to methylate receptors in an even fashion, rather than sequentially modifying a single receptor until it is fully (de)methylated. Since brachiation enables some randomization of enzyme position between methylation events, it should lead to a more distributive methylation process.

| Numerical model | Features |
|---|---|
| **M1** | Reference model; assistance neighborhoods and enzyme brachiation; activity-dependent binding kinetics; MWC receptor cooperativity. |
| **M2** | Derived from **M1**; no assistance neighborhoods or enzyme brachiation. |
| **M3** | Derived from **M1**; less efficient brachiation relative to **M1**. |
| **B1** | No enzyme tethering or lattice structure; activity-dependent binding kinetics; MWC receptor cooperativity. |
| **B2** | Derived from **B1** by increasing enzyme-receptor affinities. |

**Table 1**: Summary of numerical models.

To investigate how distributivity affects adaptation we compared our reference model **M1**, which includes assistance neighborhoods and brachiation, to a model in which the binding of tethered enzymes to the modification sites of neighboring receptors (and modification site-bound enzymes to neighboring tethers) is not allowed, denoted **M2** (Table 1). Disabling these reactions both removes assistance neighborhoods and prevents enzyme brachiation. As a result, methylation is more processive. In this scheme, an enzyme remains bound to and modifies only a single receptor during its residence time in the cluster. This scheme increases the probability that CheR and CheB will become bound to receptors with high or low methylation levels, respectively. Consequently, enzymes will tend to have low affinity for their local modification sites and modification will proceed in an inefficient manner compared to a distributive scheme. In **M2**, adaptation to both small (5 μM) and large (1 mM) steps of the attractant methyl-aspartate becomes much slower (Fig 2B, light gray) than in the reference model **M1** (Fig. 2B, black). Precise adaptation is also severely compromised for the large stimulus.

We also consider the case in which enzyme brachiation is made less efficient, but adaptational assistance is not eliminated. To examine this intermediate model (**M3**), we decreased the unbinding rates from the tether $d^t_{r,b}$ relative to **M1**. As a result, more methylation events occur before an enzyme moves on the lattice. This leads to less efficient brachiation than in **M1** but preserves assistance neighborhoods. As a result, adaptation to the large stimulus is less precise compared to **M1** but more precise than **M2** (Fig. 2B).

The picture that emerges is that the distributivity of the modification process is an important determinant of the precision of adaptation. Adaptational assistance and enzyme brachiation increase the distributivity of modification and lead to more precise adaptation in our model of the full receptor lattice. This result extends previous findings that the ability of tethered CheR and CheB to modify several receptors within an assistance neighborhood is necessary for precise adaptation within a single MWC complex [28,38]. In our simulations, as in these previous studies, increasing the distributivity of receptor methylation reduces the time CheR and CheB spend bound to highly methylated and demethylated receptors, respectively. Consequently, the methylation rate in distributive models is largely independent of the methylation levels of individual receptors, resulting in more precise adaptation. Additionally, (de)methylation rates are higher than in the more processive schemes because the enzymes spend less time interacting with receptors that are already highly methylated or demethylated. Indeed, plotting the rate of methylation after the step stimulus for the three simulations depicted in Fig 2B (bottom panel) indicates that it is highest in the most distributive model **M1** (Fig. S7 and Text S1).

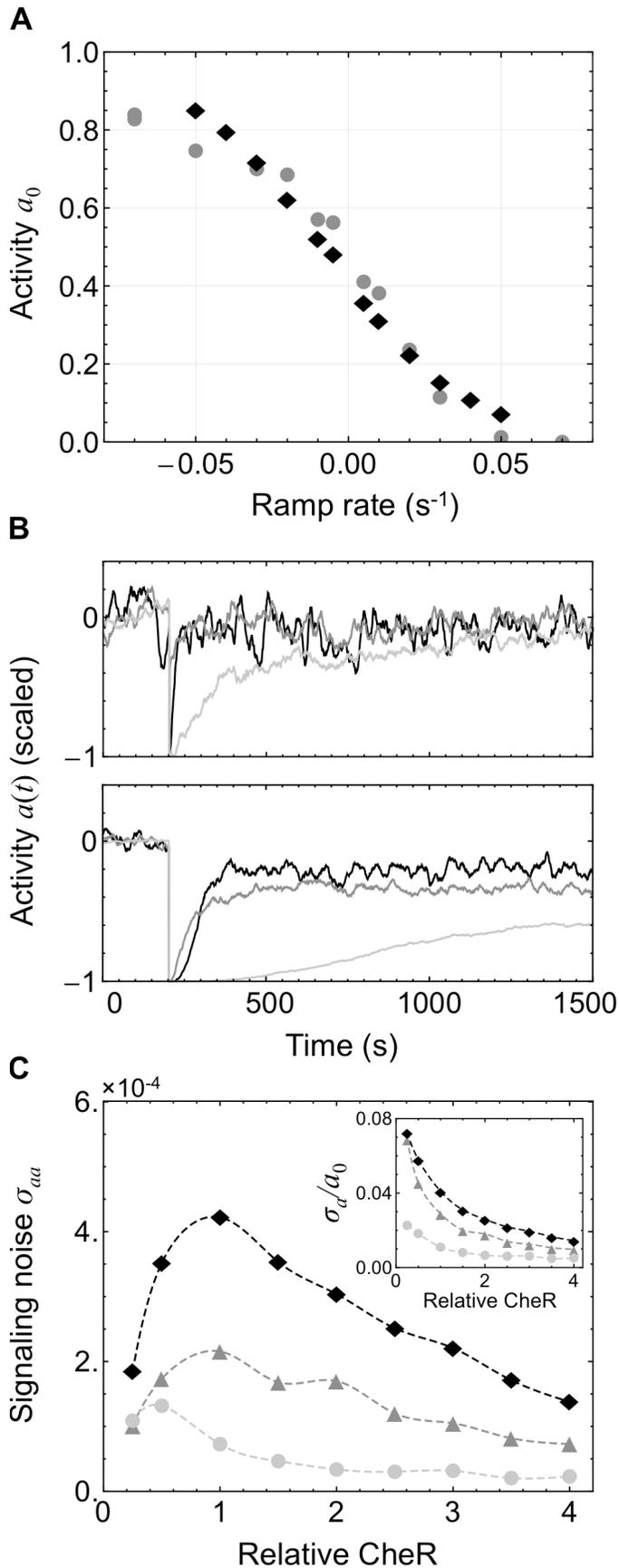

**Fig. 2**: Processive receptor methylation compromises adaptation and decreases signaling noise. Compared are three simulated models of the chemotaxis adaptation system: **M1** with assistance neighborhoods and efficient brachiation (black traces), **M2** with no assistance neighborhoods or brachiation (light gray), and **M3** with assistance neighborhoods but inefficient brachiation (dark gray). Methylation is more processive in **M2** and **M3** than in **M1**. As processivity increases, enzymes become more localized to receptors that are already highly methylated (CheR) or demethylated (CheB), limiting their effectiveness. (A) The kinetics of **M1** were calibrated by comparison to population-level measurements (gray) [43]. The model was exposed to simulated time-varying exponential ramps of methyl-aspartate and the resulting steady-state activity $a_0$ recorded (black). (B) Response to small (5 μM) and large (1 mM) MeAsp step stimulus at applied at $t = 200$ s as measured by receptor activity $a(t)$. While all models adapt to the small stimulus (top), they fail to adapt precisely to the large stimulus (bottom). For the large stimulus, higher processivity leads to less precise adaptation with **M1** performing best and **M2** worst. Activities have been scaled and recentered with steady-state values at 0. (C) Increasing processivity also decreases the magnitude of fluctuations in $a(t)$ in the adapted state around the mean value $a_0$. Plotted is the variance $\sigma_{aa}$ of $a(t)$ and the noise relative to the mean output $\sigma_a/a_0$ (inset) for different expression levels of the enzyme CheR. Fluctuations are largest in **M1** and smallest in model **M2**.

**Distributive methylation leads to large steady-state fluctuations**

Experiments and modeling efforts strongly suggest that the adaptation mechanism of the bacterial chemotaxis system introduces slow spontaneous fluctuations in the activity of the receptor-kinase complex with a standard deviation of ~5-10% of the mean [33,39-42,48,49]. These fluctuations are thought to lead to long-tailed distributions of run durations [39,50] and may enhance navigation in shallow gradients and exploration [30,32,33,35,39]. Since distributivity affects the kinetics of adaptation, it is also likely to affect the spontaneous fluctuations of the system. Fig. 2C compares the level of fluctuation in receptor activity about the unstimulated steady-state level for each model at different expression levels of CheR. The model **M1** exhibits fluctuations of the same order as those measured experimentally, particularly at low CheR levels for which the standard deviation $\sigma_a$ of fluctuations exceeds 7% of the mean activity $a_0$. Notably, the magnitude of this noise is reduced when receptor modification is made less distributive in models **M2** and **M3**. These results suggest that the features required for successful adaptation, assistance neighborhoods and brachiation, also lead to experimentally observed levels of signaling noise. The mechanism underlying these relations will be discussed in a later section with insights provided by an analytical model.

Cells within an isogenic wild-type population are known to exhibit significant cell-to-cell variability in the level of signaling noise [33,39-41]. To what extent does this variability arise from cell-to-cell variability in the expression levels of the chemotaxis proteins? Our simulations of the model **M1** indicate that the level of signaling noise is sensitive to the relative amounts of CheR and CheB in the cell (Fig. 2C). However, the multicistronic organization of *cheR* and *cheB* on the chromosome ensures that the ratio of CheR to CheB is approximately conserved in each cell within a wild-type population due to cotranscription [44,51]. Therefore variability in signaling noise levels must arise largely from correlated variation in the expression levels of the chemotaxis proteins. Using our stochastic simulation of enzyme dynamics on the receptor lattice (**M1**), we investigated the effects of covarying the number of CheR, CheB and chemoreceptors. We sampled cells from across a population in which CheR, CheB and chemoreceptor counts all vary according to a log-normal distribution (Fig. S5) obtained from measurements of CheY-YFP levels expressed from the native chromosomal locus [44]. Mean protein expression levels were set according to immunoblotting measurements [52]. To study only the effects of concerted variation in protein levels, we ignored intrinsic noise, thereby preserving the ratio of CheR/CheB/receptors. We found that the level of signaling noise varies widely between each sampled cell, between 3 and 10% of the mean (Fig. 3A). This degree of variation in signaling noise levels agrees well with measurements performed across a wild-type population [40,41]. Additionally, we found that cells with low expression levels of the chemotaxis proteins are predicted to exhibit the large fluctuations, ~10% of the mean level. Consequently, we expect cells with high levels of signaling noise to be present even in populations across which the CheR to CheB ratio is maintained at the single cell level.

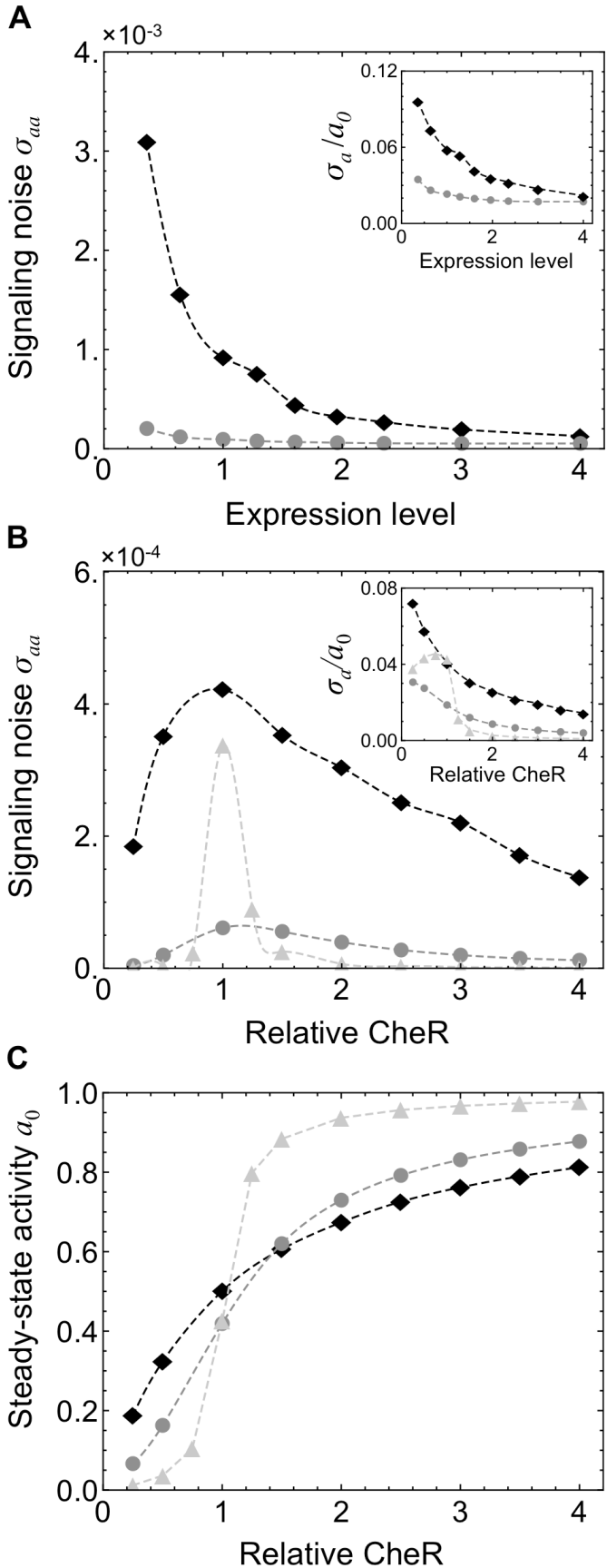

**Fig. 3**: Spontaneous output of the bacterial chemotaxis system. Results are from stochastic simulations of a chemotaxis model **M1** with a hexagonal receptor lattice and explicit enzyme tethering and the model **B1** with no tethering or lattice structure. (A) We sampled representative cells from a population in which the ratio CheR/CheB/chemoreceptors is maintained but the overall expression level varies. Stochastic simulation of model **M1** (black) predicts that some cells in this population will exhibit especially large fluctuations $\sigma_a/a_0 \sim 10\%$. The magnitude of fluctuations increases sharply as the level of protein expression decreases. Noise levels in **M1** are significantly larger than in **B1** (gray) at all expression levels. The horizontal axis is normalized by the most common expression level. (B) The variance $\sigma_{aa}$ of fluctuations in receptor activity is shown as CheR is varied while all other proteins are expressed at their mean levels. The variance $\sigma_{aa}$ is significantly greater in **M1** (black, diamonds) than in **B1** (gray, circles). The model **M1** produces exceeding 7% of the mean level (black, inset), while noise in **B1** remains less than ~3% (gray, inset). The noise was increased in **B2** by increasing the enzyme-receptor affinities tenfold (light gray) relative to **B1**. (C) **M1** and the (black, diamonds) and **B1** (gray, circles) also exhibit similar dependence of the mean receptor activity at steady state $a_0$ on CheR count. The model **B2** with higher enzyme-receptor affinities exhibits highly ultrasensitive dependence on the CheR count (light gray).

**High levels of signaling noise occur around a robust adapted level**

In previous models of the chemotaxis system in which enzyme localization is not considered, the slow, spontaneous fluctuations in the activity of the system were traced back to the ultrasensitive nature of the methylation and demethylation reactions, which were assumed to operate near saturation [30]. This mechanism, however, is insufficient to explain the large magnitude of the noise observed experimentally in individual cells. Indeed, using a stochastic simulation of a recent representative analytical model (Model **B1**) in which the authors calibrated the rates of methylation-demethylation using direct measurements of the average response of the receptor activity to ramps of attractant [43], we observe at most 2-3% relative noise for the individual cell (Fig. 3B). The model **B1** incorporates activity-dependent binding of the enzymes to the modification sites, but does not consider any aspects of enzyme localization via tether binding (Table 1). Additionally, while this model includes cooperative receptor-receptor interactions using a MWC model, given by Eq. (13) as for **M1**, it considers neither the geometry of the receptor cluster nor the resulting features of adaptational assistance and enzyme brachiation. Higher noise levels can be obtained in this model by increasing the enzyme-substrate affinities tenfold (Model **B2**). These higher affinities, however, result in a steady-state activity that is ultrasensitive to total enzyme counts (Fig. 3C, light gray). In this case the addition or subtraction of only a few adaptation enzymes in the cell is sufficient to switch the system between the fully active and fully inactive states. This scenario is biologically unacceptable since small fluctuations in gene expression across a population would lead to large numbers of non-functional cells with either fully active or inactive receptors at steady state. Parameter values for models **B1** and **B2** are given in Tables S4 and S6.

Interestingly, in our model accounting for the localization of enzymes to the receptor cluster, large fluctuations around the steady state activity are present even though the mean activity remains relatively robust to changes in enzyme counts. Fig. 3B shows the dependence of the steady-state fluctuations in **M1** on total CheR count with all other parameters fixed. **M1** exhibits activity fluctuations that exceed 7% of the mean value $a_0$ for low CheR counts and are significantly larger than those of the model **B1** for all CheR values. While the noise level is high, the mean receptor activity at steady state, $a_0$, is only modestly sensitive to changes in the total CheR count (Fig. 3C, black). The specific features enabling the coexistence of large fluctuations with a robust steady state are discussed in a later section with reference to an analytical model.

Finally, we compare the noise levels predicted by the models **M1** and **B1** across a cell population. When cell-to-cell variability in receptor and enzyme counts is taken into account we observe that **B1**, which does not account for receptor clustering or enzyme localization, exhibits insufficiently large fluctuations ($\sigma_a/a_0 < 4\%$) across the entirety of the population (Fig 3A). In contrast, **M1** exhibits levels of noise similar to those measured experimentally [33,40,41], as discussed in the previous section.

**Mean-field model with distributive receptor methylation and precise adaptation**

To investigate the mechanisms underlying our numerical results, we constructed an approximate model that can be solved analytically. Here we provide a mathematical derivation of the model. Analysis of the adaptation mechanism using this model is provided in the next section.

At the heart of this model is a covalent modification scheme that describes the kinetics of receptor methylation by CheR and CheB, similar in form to previous models [12,25,30,53,54]. In order to modify the receptors, however, we require that CheR and CheB be localized to the receptor cluster by being bound to the tether site. In this treatment, CheR may exist in three states: free and dispersed in the cytoplasmic bulk ($R$), bound only to the tethering site of a receptor ($R^*$), and bound to both the tether site and modification site of receptors ($\overline{R^*T}$). The notation for the states ($B_p, B_p^*, \overline{B_p^*T}$) of phosphorylated CheB is analogous. Unphosphorylated CheB is assumed not to interact with the receptors and therefore only exists in the bulk ($B$). For simplicity, we assume that enzymes in the bulk always bind the higher-affinity tether sites on the receptors prior to binding the modification sites. Since the model includes reactions occurring in multiple volumes and will later be used for stochastic calculations, all molecular species below are quantified by number rather than concentration. Therefore, the binding rates as written implicitly include a factor of the inverse of the reaction volume. In the model, active receptor complexes phosphorylate CheB at a rate $a_p$ and CheB autodephosphorylates at rate $d_p$, leading to $dB_p/dt = a_p T_{Tot} aB - d_p B_p$, which we take to be in the steady state, yielding $B_p = a_p T_{Tot} aB/d_p$. We assume that only bulk CheB ($B$, $B_p$) participates in the phosphorylation reactions.

Defining $R_{Tot}^* = R^* + \overline{R^*T}$ and $B_{p,Tot}^* = B_p^* + \overline{B_p^*T}$ as the total number of tether-bound CheR and CheB-P, the dynamics of enzymes in the bulk binding to the tether site is modeled by

$$\frac{d}{dt} R_{Tot}^* = a_r^t T_{Tot} R - d_r^t R^* \qquad (1)$$

$$\frac{d}{dt} B^*_{p,Tot} = a^t_b T_{Tot} B_p - d^t_b B^*_p. \quad (2)$$

Here $(a^t_r, a^t_b)$ denote the rates of cytoplasmic enzymes binding the tether site and $(d^t_r, d^t_b)$ denote the rates of enzymes bound only to the tether unbinding the tether and dispersing into the bulk. Since the number of tether sites greatly exceeds the number of CheR and CheB [52], we assume it to be constant and equal to the total number of receptors $T_{Tot}$. Enzymes bound to the tether may bind the modification site according to

$$\frac{d}{dt} \overline{R^*T} = a^m_{r*}(1-a)R^* - \left(d^m_r + k_r\right)\overline{R^*T} \quad (3)$$

$$\frac{d}{dt} \overline{B^*_p T} = a^m_{b*} a B^*_p - \left(d^m_b + k_b\right)\overline{B^*_p T}, \quad (4)$$

in which $(a^m_{r*}, a^m_{b*})$ are the rates of a tether-bound enzyme to bind the receptor modification site, $(d^m_r, d^m_b)$ are the unbinding rates from the modification site, and $(k_r, k_b)$ are catalytic rates for demethylation and methylation of the modification site, respectively. Binding to the modification site is dependent on the activity of the receptor. Eqs. (3, 4) employ a mean-field approximation by assuming that the activity of the receptor whose modification site is to be bound is equal to the mean activity of all receptors in the cell, $a$. This assumption makes the methylation process in this model fully distributive. Therefore the mean-field model represents the limit of a single, maximally large assistance neighborhood, encompassing all receptors, or infinitely fast brachiation, in which enzymes completely randomize their position on the lattice between methylation events. Relaxing this assumption requires a more detailed analytical model, which is explored in the Supporting Text S1.

Since Eqs. (3, 4) describe a binding reaction confined to the ~5nm radius defined by the tether [46], the kinetics are fast relative to other reactions in the model (Text S1). We take $d\overline{R^*T}/dt = d\overline{B^*_p T}/dt = 0$, leading to an expression for the number of enzymes bound to both tethers and modification sites

$$\overline{R^*T} = \frac{a^m_{r*}}{d^m_r + k_r}(1-a)R^* \equiv \frac{1-a}{K_r} R^* \quad (5)$$

$$\overline{B^*_p T} = \frac{a^m_{b*}}{d^m_b + k_b} a B^*_p \equiv \frac{a}{K_b} B^*_p. \quad (6)$$

Here $K_r$ and $K_b$ are dimensionless constants analogous to Michaelis-Menten constants. The rate of change of the *total* methylation level $M$ of all MWC complexes in the system (the total number of methylated receptor sites across all receptors in the cell) is

$$\frac{dM}{dt} = k_r \overline{R^*T} - k_b \overline{B^*_p T} = \frac{k_r}{K_r}(1-a)R^* - \frac{k_b}{K_b} a B^*_p. \quad (7)$$

Using Eqs. (5-7), we write the equation describing changes in average methylation level per 2N-receptor MWC complex, $m = M(2N/T_{Tot})$, in the form familiar from the Goldbeter-Koshland system [12,30,54]

$$\frac{dm}{dt} = \frac{2N}{T_{Tot}} \left[ \frac{k_r R^*_{Tot}(1-a)}{K_r + 1 - a} - \frac{k_b B^*_{p,Tot} a}{K_b + a} \right] + \eta_m. \quad (8)$$

The tether-binding reactions Eqs. (1, 2) may be rewritten in terms of $R^*_{Tot}$ and $B^*_{p,Tot}$ as

$$\frac{d}{dt} R^*_{Tot} = a^t_r T_{Tot} R - d^t_r \frac{K_r}{K_r + 1 - a} R^*_{Tot} + \eta_r \quad (9)$$

$$\frac{d}{dt} B^*_{p,Tot} = a^t_b T_{Tot} B_p - d^t_b \frac{K_b}{K_b + a} B^*_{p,Tot} + \eta_b \quad (10)$$

with an activity-dependent unbinding step. To include variation around the mean, Langevin sources ($\eta_m$, $\eta_r$, $\eta_b$) have been added with magnitudes evaluated using the linear noise approximation (Text S1) [55,56]. The instantaneous output of the system is the fraction of active receptors $a(t) = a[m(t), L(t)]$ with $a$ given by a MWC model, Eq. (13), for some external stimulus $L(t)$ (Methods) [22,23,43]. The noise statistics of the output $a(t)$ at steady state are calculated by linearizing the model and solving it as a multivariate Ornstein-Uhlenbeck process (Methods and Text S1) [57,58]. Parameter values for the analytical model (Tables S1 and S4) were taken to be consistent with those of the stochastic simulation **M1**.

Two important features can be noted from the form of Eqs. (8-10). First, Eqs. (9) and (10) emphasize that unbinding from the receptor lattice is a two-step process. Since CheR has higher affinity for the modification site as activity decreases, the overall rate of CheR unbinding the lattice and returning to the bulk decreases accordingly. Additionally, a smaller value of $K_r$, which denotes higher affinity of the localized enzyme for the modification site, leads to slower overall rates of unbinding. The argument for CheB-P unbinding is analogous. Second, since Eq. (8) depends only on the mean activity of the system and not on methylation or stimulus levels, the analytical model exhibits precise adaptation. This property follows from the mean field assumption or, equivalently, the assumption of fully distributive kinetics.

Using this analytical model, we next examine the mechanisms underlying the key observations made using numerical simulations and argue that: (1) large fluctuations in receptor activity are primarily due to noise in localized enzyme counts amplified by a methylation process ultrasensitive to these counts; (2) a distributive methylation scheme increases signaling noise by increasing the ultrasensitivity of this process; (3) the localized enzymes work at saturation without causing the mean activity to be ultrasensitive with respect to total enzyme expression levels. This result contrasts with the covalent modification scheme studied by Goldbeter and Koshland [12].

## High levels of signaling noise arise from fluctuations in localized enzyme counts amplified by saturated methylation kinetics

The analytical model derived above predicts large fluctuations in receptor activity (Fig. 4A, black), similar to those predicted by the stochastic simulation **M1**. This level of signaling noise is significantly higher at all CheR levels than the level predicted when enzyme localization is not taken into account (Fig. 4A, gray; analytical version of model **B1** [43]). The high level of intracellular signaling noise in this system arises from three key features.

First, since the total numbers of CheR and CheB are small [52], the relative variation in the number of localized enzymes due to Poisson statistics is large. The overall rates of methylation and demethylation are therefore highly variable in time. Second, these fluctuations in localized enzyme counts occur at sufficiently slow time scales [36] to not be filtered out by the methylation process. The possibility of slow fluctuations in the number of tethered enzymes leading to increased fluctuation in receptor activity was previously noted using a model of a single MWC complex [38]. Third, the interaction between the localized enzymes and the substrate occurs at saturation. Since the binding of the localized enzymes to the receptor modification site is activity-dependent, this interaction takes the same form as the covalent modification system studied by Goldbeter and Koshland [12], as can be seen from Eq. (8). Therefore we may analyze the localized enzyme-receptor interaction in the same terms. Since a localized enzyme is confined to the tether radius, the effective local substrate concentration is high and binding to the modification site proceeds at a fast rate. Therefore, $K_r$, $K_b \ll 1$ and, following Goldbeter and Koshland, the steady-state output $a_0$ has ultrasensitive dependence on the ratio of *localized* CheR to CheB-P (Fig. 4B, steep curve). This steep relationship suggests that the output of the system is in general highly susceptible to changes in the ratio of localized CheR to CheB-P and, consequently, fluctuations in this ratio are the primary source of noise in the output. In the limit in which methylation is fast relative to enzyme localization, $dm/dt \sim 0$, Eq. (8) yields $a = a\left(R^*_{Tot}/B^*_{p,Tot}\right)$. In this limit, receptor activity is a function of only the ratio of the localized adaptation enzymes, corresponding to the steep curve of Fig. 4B. Likewise, the variance in receptor activity becomes $\sigma_{aa} = \left[da_0/d\left(R^*_{Tot}/B^*_{p,Tot}\right)\right]^2 \text{var}\left(R^*_{Tot}/B^*_{p,Tot}\right)$. Therefore when the catalytic step is fast relative to enzyme localization, fluctuations in the localized enzyme ratio are amplified by exactly this steep curve. This limit case is relevant for understanding the behavior of our analytic and numerical models, in which the rates of enzyme localization are slow relative to all other rates in the system.

We may also show that fluctuations in the number of localized CheR and CheB are the dominant noise sources in the system without assuming $dm/dt \sim 0$. To illustrate this point, we use the analytical model to decompose the total variance $\sigma_{aa}$ of the receptor activity into a sum of three terms, each plotted in the inset of Fig. 4B:

$$\sigma_{aa} = \sigma_{aa,r} + \sigma_{aa,b} + \sigma_{aa,m}, \qquad (11)$$

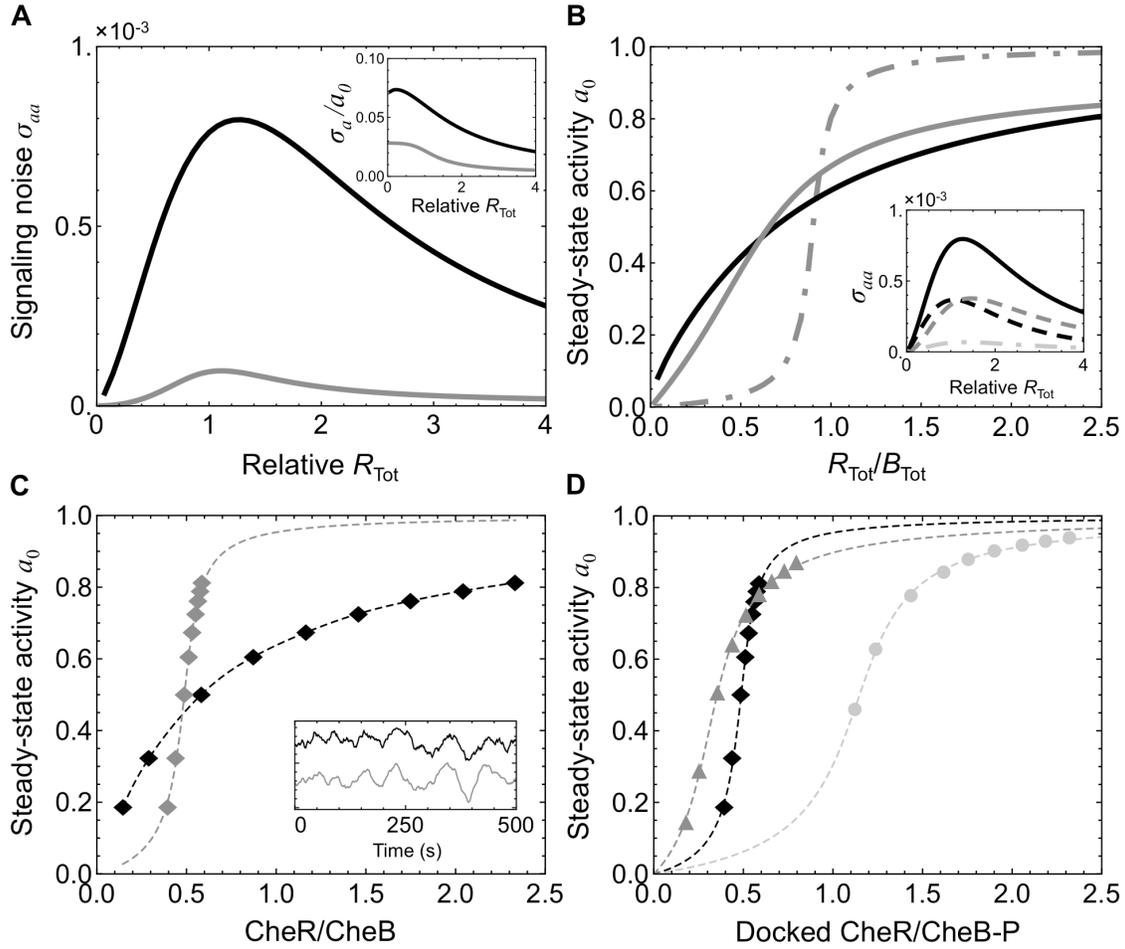

**Fig. 4**: Large fluctuations arise from the saturated kinetics of localized enzymes. (A) Variance of receptor activity $\sigma_{aa}$ at steady state is significantly larger for the analytical model with localization (black) than without localization (gray; analytical version of model **B1**) for all values of total CheR $R_{Tot}$. The analytical model with localization (inset, black) exhibits signaling noise with $\sigma_a/a_0$ up to ~7% while noise in the model with no localization (analytical version of **B1**) remains at or below 3% of the mean output (inset, gray). (B) Mean receptor activity $a_0$ at steady state as a function of CheR to CheB ratio. When plotted as a function of the *total* CheR to *total* CheB ratio, $a_0$ exhibits a similar relatively robust profile for both the analytical model with localization (black) and without localization (gray; analytical version of **B1**). In contrast the mean receptor activity is ultrasensitive to the ratio of the *localized* CheR to *localized* CheB-P counts (gray, dot-dashed), $R^*_{Tot}/B^*_{p,Tot}$. (Inset) Variance in receptor activity $\sigma_{aa}$ (black, solid) decomposed into components due to fluctuation in localized CheR (black, dashed), localized CheB (gray, dashed), and small intrinsic fluctuations in the methylation rates (gray, dot-dashed) as in Eq. (11). All quantities are plotted as functions of relative $R_{Tot}$. (C) In the stochastic simulation of **M1**, steady-state activity $a_0$ also has ultrasensitive dependence on the ratio of tethered CheR/CheB-P (gray), despite the weak dependence on total CheR/CheB (black). (Inset) 500 s simulation trace of instantaneous mean receptor activity $a(t)$ (black) and instantaneous localized CheR/CheB-P (gray), smoothed with a 30 s sliding window average. (D) Comparison of the dependence of $a_0$ on localized CheR/CheB-P for the simulated models **M1** (black), **M2** (light gray), and **M3** (dark gray) from Fig. 2. This dependence is significantly weaker for the more processive models.

fluctuations due to the number of localized CheR ($\sigma_{aa,r}$), those due to number of localized CheB-P ($\sigma_{aa,b}$), and fluctuations due to intrinsic variability in the methylation and demethylation rates ($\sigma_{aa,m}$). Each contribution $\sigma_{aa,i}$ depends linearly on the intensity of the corresponding noise source $\eta_i$ in Eqs. (8-10), $\sigma_{aa,i} \propto \langle \eta_i^2 \rangle$. The magnitude of the third term $\sigma_{aa,m}$ is comparable to the total noise predicted by models without enzyme localization. Fig. 4B (inset) shows that the first two terms on the right hand side of Eq. (11) dominate to the exclusion of the third, confirming that variability in localized CheR and CheB-P is the dominant source of the large fluctuations in receptor activity.

This same mechanism underlies the observed large fluctuations in the stochastic simulation of the model **M1**, considered previously. Fig. 4C shows mean activity $a_0$ versus the ratio of mean localized CheR to mean localized CheB-P obtained from simulation by varying only the total CheR count. As in the analytical model, this relationship is highly ultrasensitive. To illustrate the dependence between fluctuations in the localized enzyme ratio and fluctuations in receptor activity, the inset of Fig. 4C displays 500s time traces of receptor activity and the ratio of localized CheR to localized CheB-P taken from simulation. The correlation between the two series is apparent and consistent with activity fluctuations arising from variability in the number of tethered enzymes.

In summary, clustering of the receptors leads to a high density of modification sites for the enzymes localized at the cluster. This results in saturated ultrasensitive kinetics of the covalent modification reactions, which strongly amplify the noise due to the slow exchange of enzymes between the cluster and the bulk.

**Relation between distributive receptor modification and high levels of signaling noise**

In the analytical model, large fluctuations in receptor activity result from the high affinity of localized enzymes for the modification site. Since all receptors in the analytical model are assumed to have the same activity, this affinity is entirely characterized by the small values of the constants $K_r$ and $K_b$. In the numerical models, in contrast, the binding of enzymes to individual receptor dimers within MWC complexes of varying levels of activity is explicitly simulated. Consequently, the affinity of the enzymes for the modification site depends not just on the values of $K_r$ and $K_b$ (as derived from the binding, unbinding, and catalytic rates in the simulation), but also on the distribution of CheR and CheB within complexes of different activities. If enzymes tend to become localized within regions of the cluster for which they have low binding affinity (*e.g.*, CheR within a highly methylated region), we expect the ultrasensitive dependence of activity on the ratio of localized enzymes (Fig. 4C) to be reduced. This effect may be thought of as increasing the effective values of $K_r$ and $K_b$.

Adaptational assistance and brachiation mitigate this effect to some extent by enabling localized enzymes to sample a number of receptors during their residence time in the cluster. A higher rate of sampling indicates that a given enzyme samples a larger fraction of the cluster between subsequent methylation events and therefore corresponds to more distributive methylation kinetics. A potentially analogous situation has been studied theoretically for a MAP kinase cascade [13]. In this system, slowly diffusing enzymes tended to rebind the same substrate molecule multiple times, leading to a processive modification scheme. Faster diffusion enabled the enzymes to randomize their positions between modification events, corresponding to distributive modification. In the MAP kinase study, faster diffusion led to an ultrasensitive dependence of the output on enzyme levels. Is a similar mechanism at work in the chemoreceptor cluster?

For our numerical models, we quantified the rates at which enzymes sampled different, unique receptors within the cluster and found that this rate was between 4 and 13-fold smaller for the more processive models **M2** and **M3** than for the reference model **M1** (Table S7). Accordingly, the steady-state activity in the more processive models **M2** and **M3** is also less dependent on the ratio of localized CheR to CheB-P than in **M1** (Fig. 4D). Since this relationship effectively amplifies fluctuations in the ratio of localized enzymes, this decreased steepness leads to lower signaling noise levels in these more processive models, as seen previously (Fig. 2C). For further details regarding the comparison between simulations and the analytical model, see Supporting Text S1. We conclude that a distributive methylation scheme leads to higher signaling noise levels by increasing the overall affinity of the localized enzymes for the modification site substrate.

**Localized enzymes may work at saturation without compromising robustness to cell-to-cell variability in total enzyme expression levels**

The mean steady-state activity for the analytical model with enzyme localization is plotted in Fig. 4B as a function of the ratio of both localized and total (across the entire cell) adaptation enzymes, $R^*_{Tot}/B^*_{p,Tot}$ and $R_{Tot}/B_{Tot}$. While the activity is highly ultrasensitive with respect to the localized enzyme ratio, its sensitivity to the total enzyme ratio is significantly less and comparable to the model **B1**. Therefore, the mean steady-state activity of the system $a_0$ is robust to changes in the total CheR to CheB ratio caused by noisy gene expression. This result is somewhat surprising because in the classic covalent modification system studied by Goldbeter and Koshland [12], saturated enzyme-substrate interactions always lead to a steady-state activity that is ultrasensitive to the total CheR to CheB ratio.

In Eq. (8), which we may analyze in the same manner as the Goldbeter-Koshland system, the sensitivity of the steady-state activity $a_0$ with respect to the ratio of localized CheR to CheB is determined solely by the constants ($K_r$, $K_b$) that characterize the probability that a localized enzyme will be bound to a modification site. Small values of these constants lead to saturated kinetics and ultrasensitivity of the steady-state activity to the ratio of *localized* CheR to CheB. Our model differs from the Goldbeter-Koshland system, however, in that in our model these constants only partially determine the sensitivity of $a_0$ to the ratio of *total* CheR to CheB. The sensitivity of the system to the total enzyme ratio is also determined by the rates at which cytoplasmic enzymes localize to the cluster and at which localized enzymes return to the bulk. Since the rates $\left(a^t_r, a^t_b\right)$ at which enzymes localize to the cluster are slow [36], the effective affinities of the enzymes for the modification sites are reduced even though the affinities of enzymes already localized at the cluster are high.

The steady-state solutions to Eqs. (8-10) quantify how the mean steady-state activity depends on the total enzyme counts $R_{Tot}$ and $B_{Tot}$. Solving Eqs. (9) and (10) for the localized enzyme counts $R^*_{Tot}$ and $B^*_{p,Tot}$ and inserting the results into Eq. (8), we obtain

$$\frac{dm}{dt} = \frac{k_r R_{Tot}(1-a)}{K_r\left(1 + d^t_r/a^t_r T_{Tot}\right) + 1 - a} - \frac{k_b B_{Tot} a}{K_b\left[1 + d^t_b/a^t_b T_{Tot}\left(1 + d_p/a_p T_{Tot} a\right)\right] + a} = 0. \quad (12)$$

Eq. (12) is also of the Goldbeter-Koshland form which indicates that the steepness of the steady-state activity as a function of the total CheR to CheB ratio is determined by the effective inverse affinities $K'_r = K_r\left(1 + d^t_r/a^t_r T_{Tot}\right)$ and $K'_b(a) = K_b\left[1 + d^t_b/a^t_b T_{Tot}\left(1 + d_p/a_p T_{Tot} a\right)\right]$. Values of $K'_{r,b} \ll 1$ lead to ultrasensitivity of the steady-state activity with respect to the ratio $R_{Tot}/B_{Tot}$. For the steady-state activity to be considered robust, we require $K'_{r,b} \sim 1$. From this condition, we can see that the steady-state $a_0$ can be robust even if the affinity of the localized enzymes for the modification site is extremely high, $(K_r, K_b) \ll 1$. This will be the case if the rates $a^t_{r,b}$ of enzymes in the bulk to reach the cluster and bind the tether are sufficiently small relative to the unbinding rates $d^t_{r,b}$, effectively compensating for the small $(K_r, K_b)$ and leading to $K'_{r,b} \sim K_{r,b} d^t_{r,b}/a^t_{r,b} T_{Tot} \sim 1$.

To discuss the robustness of the bacterial chemotaxis system, we note three key considerations. First, we estimate that $K_r$, $K_b \ll 1$ due to the fast rate of the highly localized enzymes binding the modification site (Text S1). Second, we note that the CheB-P feedback loop is not by itself sufficient to make the steady-state robust to the total enzyme ratio. While the term due to the feedback loop in $K'_b$, $1 + d_p/a_p T_{Tot} a$, is greater than 1 and therefore confers some degree of robustness, for typical values of activity, $a \sim 0.2$ or greater, the term is only of order 1 and therefore not sufficient to compensate for small $K_b$. Robustness therefore likely arises from the slow kinetics of tether binding. The final consideration is that measurements [36] indicate that the number of cytoplasmic and localized enzymes are comparable and therefore that the forward and reverse rates of Eqs. (9) and (10) are roughly equal. This condition not only leads to comparable numbers of localized and cytoplasmic enzymes, but also indicates that the rates of tether binding and unbinding fall in the regime in which the steady-state activity is robust to the total number of enzymes. Specifically, for CheR, requiring the forward and backward rates of Eq. (9) to be comparable yields $a^t_r T_{Tot} \sim d^t_r K_r/(K_r + 1 - a) \sim d^t_r K_r/(1-a)$, leading to $K_r d^t_r/a^t_r T_{Tot} \sim K'_r \sim 1$ for typical values of $a$ (0.3-0.5) [43]. The argument for CheB is analogous. Satisfying this constraint therefore leads not only to both comparable numbers of localized and cytoplasmic enzymes, but also to a steady-state activity that is robust to the total enzyme ratio. In this manner, the steady-state of the bacterial chemotaxis system can remain robust even when the localized enzymes operate at saturation.

## Discussion

Chemotactic bacteria are able to navigate chemical gradients with strengths ranging over five orders of magnitude [19]. This remarkable capability results from the capacity of the system to amplify small input signals while adapting to a wide range of concentrations of persistent stimulus. The cooperative receptor-receptor interactions that amplify input signals are facilitated by the formation of large receptor clusters, structures that are strongly conserved across bacterial species [5]. Adaptation to stimulus requires the efficient recruitment of cytoplasmic enzymes to these clusters, which is achieved through the presence of a high-affinity enzyme-tethering site on most receptors. These tethers, together with the dense structure of the receptor lattice, give rise to assistance neighborhoods [27] and possibly enzyme brachiation [37]. These features increase the distributivity of methylation, decreasing the likelihood that enzymes become localized in neighborhoods within which they have low binding affinity and therefore act inefficiently.

Building on previous work that showed assistance neighborhoods were necessary for precise adaptation in a single strongly coupled signaling complex [28,38], we found that assistance neighborhoods and enzyme brachiation contributed to precise adaptation to stimulus. We further linked distributive methylation to the presence of signaling noise in the output and showed how high signaling noise may coexist with a mean level of receptor activity that is robust to changes in the ratio of the adaptation enzymes. This ratio is not exactly conserved across populations. Consequently, if the mean activity were not sufficiently robust, the ultrasensitivity of the flagellar motor [59,60] would lead to a significant fraction of nonfunctional cells permanently in the running or tumbling state. This robustness to the ratio of adaptation enzymes occurs even though the localized enzymes work in the saturated regime. This scheme is not possible for the simpler covalent modification system studied by Goldbeter and Koshland, in which saturated enzyme kinetics always corresponds to ultrasensitivity to the enzyme ratio.

The mechanism described here is not necessarily restricted solely to the bacterial chemotaxis system. The analytical model presented in this study describes generally an extension of the Goldbeter-Koshland [12] motif in which enzymes transition between active and inactive states, whether by localization to the substrate prior to modification, as in the bacterial chemotaxis model, or by chemical activation of the enzyme. This simplified model captures the essential features underlying large fluctuations: slow enzyme activation relative to the modification rate, saturated kinetics between the activated enzyme and the substrate, and distributive modification. While the kinetics of activated enzyme and substrate may be saturated, the robustness of the system to the overall expression levels of the enzymes may be preserved if the enzyme activation (localization) rate is sufficiently small relative to the deactivation (delocalization) rate. The effects of enzyme localization and the relationship between rapid enzyme rebinding and processivity have been considered in studies of MAP kinase cascades. A recent study of the mating response in yeast [61] discusses a mechanism in which the kinase Fus3 and phosphatase Ptc1 bind a docking site on the substrate Ste5 prior to modification. Since the docked enzymes operate at saturation, the system is ultrasensitive to changes in the number of recruited enzymes, similar to the chemoreceptor-enzyme system discussed in this work. Unlike the chemotaxis system, however, yeast exploits these saturated kinetics to produce a switch-like response in the steady state. The theoretical work of Takahashi *et al*. [13] also considers the MAP kinase system, using it as a model to explore the role of enzyme diffusion in determining whether substrate modification is processive or distributive. The authors conclude that slow diffusion, which causes the enzyme to bind and phosphorylate the same substrate molecule repeatedly, can effectively convert a distributive mechanism into a processive one, reducing the sensitivity of the system. The same effect figures prominently in our model of the bacterial chemotaxis system but in the opposite regime, in which the brachiation process serves to randomize enzyme positions between methylation events.

Future studies of the bacterial chemotaxis system may further clarify the role of enzyme brachiation in adaptation. Different configurations of clustered receptors from that considered here, such as less dense clusters that have been shown to reduce cooperativity [62], or larger numbers of significantly smaller clusters [63], could hinder the ability of localized enzymes to visit a large number of unique receptors. In these cases our results suggest that signaling noise would be reduced. Interestingly, brachiation may be particularly important when considering cluster structure within local adaptation models [64]. In these models, receptors of different types respond specifically to different stimuli. Consequently, successful adaptation may depend on the ability of the adaptation enzymes to localize efficiently to responsive receptors. Brachiation may be critical for such efficient localization, particularly when considering the adaptation of low abundance receptors to their specific stimuli.

While many systems benefit from minimizing signaling noise, studies of bacterial chemotaxis have shown that noise may increase the performance of the system in sparse environments while introducing only minimal deleterious effects. In empty environments, signaling noise may lead to faster cellular exploration to locate nutrient sources more efficiently [32,33,39]. Signaling noise has also been shown theoretically to increase tracking performance in shallow gradients [32,33,35]. These results are consistent with a picture of the chemotaxis system being not purely a signal transduction system, for which minimizing noise would typically be desirable, but also a feedback system in which the output controls the sampling of the input.

## Methods

### Receptor activation

Since changes in receptor activity are effectively instantaneous relative to the slow methylation kinetics, activation of the receptor clusters is described by an equilibrium MWC model [22,23]. Clusters in the model are composed of $N = 6$ Tar homodimers. The free energy difference between the active and inactive states of the cluster is decreased by $\varepsilon_1$ per methylation level and increased by $N\log\left[(1+L/K)/(1+L/K^*)\right]$ in the presence of methyl-aspartate attractant $L$. Then the fraction of active clusters is given by

$$a(m,L) = \frac{1}{1+\exp(\varepsilon_0 - \varepsilon_1 m)\left(\frac{1+L/K}{1+L/K^*}\right)^N} \quad (13)$$

with $m$ the methylation level. Parameter values were taken from fits to dose response measurements [43] and reproduced in Table S1. In the stochastic simulation, $m$ is taken to be the methylation level of a single MWC signaling unit and $a(m, L)$ is used to calculate the activity of each MWC unit individually. In the analytical model, following Shimizu *et al*. [43], $m$ is the average methylation level per receptor cluster and $a(m, L)$ is taken to be the average activity of all receptors in the system.

**Signaling properties**

We analyze the signaling properties of the model Eqs. (8-10) by performing a perturbation analysis around the steady state. Small displacements in the numbers of chemical species **x** evolve according to the linear system of Itō stochastic differential equations

$$d\mathbf{x}(t) = A\mathbf{x}(t)dt + Bd\mathbf{W}(t) \quad (14)$$

in which $A$ is the Jacobian matrix of the system, $B$ is the diffusion matrix, and $\mathbf{W}(t)$ is the multidimensional Wiener process. By the linear noise approximation, $B^T B = S\,\mathrm{diag}(\mathbf{v})\,S^T$ with $S$ the stoichiometry matrix and $\mathbf{v}$ the propensity vector [55,56]. The system in Eq. (14) is a multivariate Ornstein-Uhlenbeck process [57]. $A$ has eigenvalues with negative real components, indicating the system relaxes to steady state after perturbation. The steady-state variance in the output of the system is obtained by solving the Lyapunov equation

$$A\sigma + \sigma A^T + B^T B = 0 \quad (15)$$

for the covariance matrix $\sigma$. Additional details of the noise calculation are presented in the Supporting Text S1.

# Acknowledgements

The authors thank Tom Shimizu, Yann Dufour, and Nicholas Frankel for helpful discussions and comments on the manuscript.

# Supporting Text S1

Adaptation dynamics in densely clustered chemoreceptors

William Pontius[1,2], Michael W. Sneddon[2,3,4], Thierry Emonet[1,2]

[1]Department of Physics, Yale University, New Haven, CT, USA

[2]Department of Molecular, Cellular, and Developmental Biology, Yale University, New Haven, CT, USA

[3]Interdepartmental Program in Computational Biology and Bioinformatics, Yale University, New Haven, CT, USA

[4]Current address: Physical Biosciences Division, Lawrence Berkeley National Lab, Berkeley, CA, USA

**NFsim implementation of the receptor lattice model**

NFsim [1] is a rule-based stochastic simulator of chemical reaction networks built on the BioNetGen language [2]. It is designed to efficiently simulate systems in which molecules may exist in large numbers of states, and in which these states affect the rates of the reactions in which molecules participate. To illustrate the problem, we consider the case of a bacterial chemoreceptor in a MWC signaling complex. The rate of CheR binding to the modification site depends on: (1) whether the receptor modification site is occupied (by CheR or CheB); (2) whether the enzyme active site is occupied (by CheR or CheB); (3) whether the enzyme is tethered to the receptor, tethered to a neighboring receptor, tethered to a non-neighboring receptor, or in the bulk; and (4) the methylation level of the signaling complex in which the receptor is located, which varies between 0 and 48 for a complex of six dimers. Accordingly, the reaction proceeds with a rate specific to each of the 3×3×4×49 = 1764 possible receptor-enzyme states. NFsim enables us to fully specify the above model with relatively few explicit reaction rules. Moreover, the speed of simulation in NFsim scales nearly independently of the number of possible states [1].

In the simulation, chemoreceptor dimers are specified by objects of the form `T(m,mc,as,teth,[loc],[hex])` in which `m` and `mc` denote the methylation level of the dimer (0 to 8) and the local MWC cluster (0 to 48), `as` and `teth` are binding sites representing the active site and tether respectively, and `[hex]` and `[loc]` are each a series of binding sites used to specify the organization of the receptor lattice. While NFsim does not support spatially resolved simulations, we can specify the neighbors of a given dimer by creating bonds between it and all of its neighboring dimers. Fig. S1A illustrates how a MWC cluster of six dimers is specified by creating bonds (blue lines) between the `[loc]` sites (blue squares) on each dimer. Fig. S1B illustrates 21 MWC clusters assembled into a hexagonal lattice by specifying bonds (red lines) between the `[hex]` sites (red squares) of neighboring dimers. All interior dimers are connected to six neighboring dimers. These bonds need not correspond to chemical bonds in the actual system; here they are a feature of the simulation language that we use to specify the lattice organization.

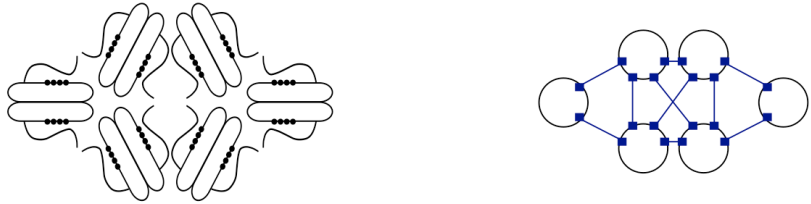

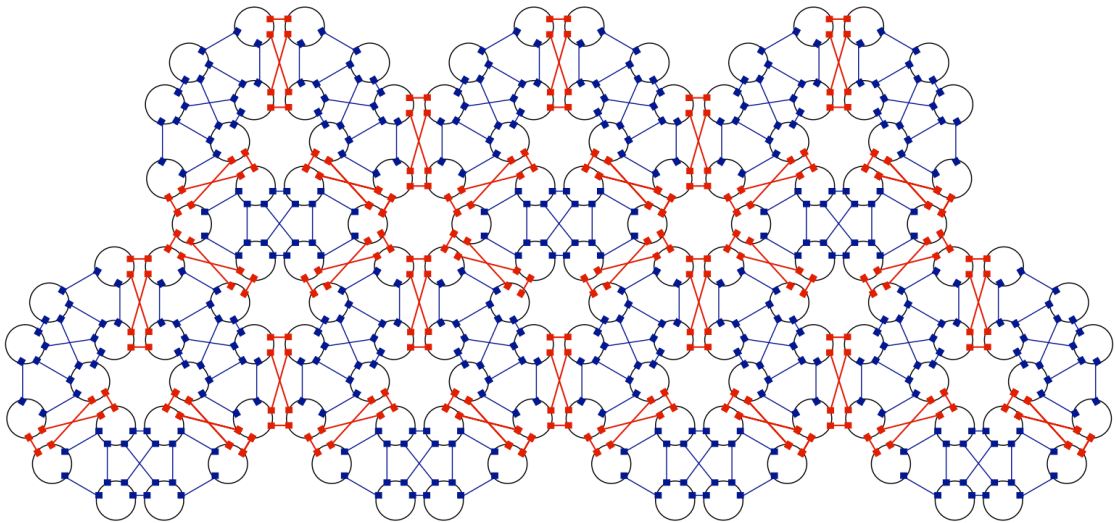

**Fig. S1**: Structuring the chemoreceptor lattice in NFsim. (A) A MWC signaling complex consisting of two trimers of dimers (left) is specified by enumerating bonds (right, blue) between a dimer and all of its neighbors within the complex. (B) The hexagonal lattice is then structured by enumerating bonds between a given dimer and all of its neighbors in other signaling complexes (red). The pictured lattice consists of 21 MWC complexes. All interior dimers have six neighbors. The basic unit of the lattice is the hexagon consisting of three signaling complexes. We model lattices of equal length and width, as specified in terms of this basic hexagonal unit.

To illustrate reaction rules, we consider the reaction in which a tethered CheR binds to the active site of a neighboring receptor. The corresponding rule is

`R(as,teth!1).T(teth!1,hex!2).T%t(as,hex!2) ->`

        `R(as!3,teth!1).T(teth!1,hex!2).T%t(as!3,hex!2).`

CheR is represented by the object R with two binding sites `as` and `teth`. The dot notation indicates that two objects are bound and the `!n` notation serves to label distinct bonds. `T%t` indicates that the reaction rate is a function of the state of the object `T`, referenced as `t` in the function argument. In this case, the reaction rate is a basal binding rate times one minus the activity of `t`, calculated in the simulation by evaluating Eq. (13) at the methylation level `mc` of `t`. Functionally defined rate laws are a key feature of NFsim that in this illustrative case save us from having to define a separate reaction for each value of `mc` and `m`.

For simplicity, we have assumed that each dimer has one modification site and one tether site. This simplification should not affect the results significantly since the number of receptors greatly exceeds the number of adaptation enzymes [3]. The simulation described in this section and in the main text was used for models **M1**, **M2**, and **M3**. Parameters for these models are given in Tables S1, S2, S6 and are discussed below.

**Implementation of models with no enzyme localization**

The models **B1** and **B2** were also simulated using NFsim. MWC signaling complexes were modeled as objects with a methylation level ranging from 0 to 48 and a modification site for enzyme binding. The activity $a$ was calculated for each signaling complex using Eq. (13). Binding of CheR and CheB to the complex was taken to be proportional to 1 - $a$ and $a$, respectively. Parameters for these models are given in Tables S1, S3, S6. **B1** is adapted from an analytical model presented in a previous study [4].

**Parameter values**

*1. Parameter values common to all models*

Values in Table S1 were taken from experimental measurements presented in previous studies. The basal protein counts ($R_{Tot}$, $B_{Tot}$, $T_{Tot}$) represent the mean counts per cell measured across a wild-type population by immunoblotting [3]. Parameters for the MWC model of Tar receptor clusters ($\varepsilon_0$, $\varepsilon_1$, $N$, $K$, $K^*$) were obtained through FRET measurements of kinase activity in response to doses of the chemoattractant methyl-aspartate [4]. The value of $\varepsilon_1$ reflects that in Eq. (13) $a(m, L)$ is written in terms of the mean methylation level $m$ per MWC signaling complex.

*2. Parameter values for numerical models of the receptor lattice and the analytical model with enzyme localization*

Parameter values for the adaptation kinetics were chosen to agree with recent *in vivo* measurements. All parameter values for the analytical model with enzyme localization (Table S4) are taken to agree with corresponding parameters in the numerical model **M1** (Table S2). The rates of localization of cytoplasmic CheR and CheB-P $a^t_{r,b}$ to the receptor cluster were taken from FRAP measurements [5]. We interpret these rates to represent enzymatic binding to the high-affinity tether sites. The rates of localization by binding the lower-affinity modification sites $a_{r,b}^m$ were taken to be slower. Since few enzymes localize through this channel, their exact values do not affect the predictions of the model significantly.

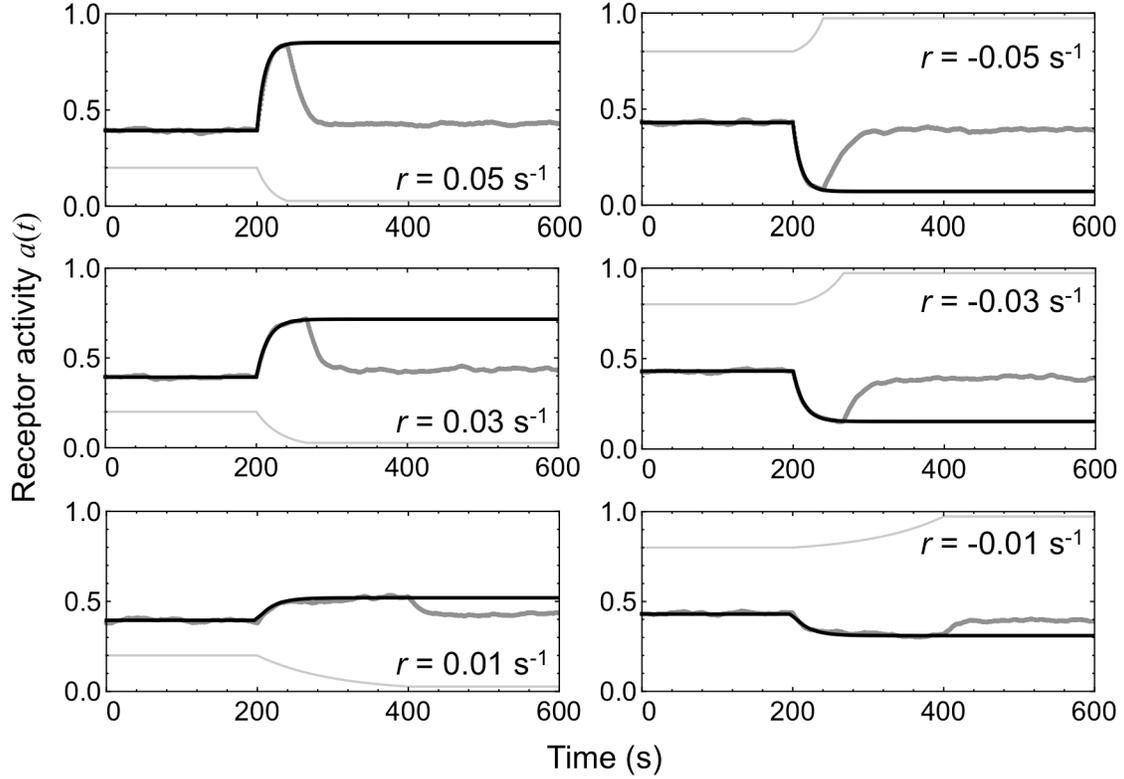

**Fig. S2**: Response of the numerical model **M1** to time-varying exponential ramps of chemoattractant. We presented the simulated cells with exponential ramps of methyl-aspartate (light gray, plotted in arbitrary units) of rate $r$ (shown in each panel) and averaged the response in receptor activity over ten trials (dark gray). For each ramp, receptor activity approached a steady-state value during stimulus, determined by exponential fits (black) to $a(t)$ and plotted in Fig. 2A of the main text. Following a recent experiment [4], the methyl-aspartate concentration ranged between 0.084 and 0.62 mM.

Key parameters for the model are $K_r$ and $K_b$ (Table S4), which characterize the affinity of tethered enzymes for the modification site and therefore the steepness of the relationship between receptor activity and the ratio of localized enzymes (Figs. 3B-D). These affinities are related to the rates in model **M1** (Table S2) via $K_{r,b} = (d_{r,b}^m + k_{r,b})/a_{r^*,b^*}^m$. The values of $K_r$ and $K_b$ correspond to the Michaelis-Menten constants for the enzyme-modification site interaction divided by the effective local concentration of the tethered enzyme. Since the tether length is on the nanometer scale, these local concentrations are high: theoretical estimates based on the tether structure vary from 0.17mM [6] to 5M [7]. Given this range of estimates, we chose the values of $K_r$ and $K_b$ conservatively (*i.e.*, to be relatively large). For example, assuming Michaelis-Menten constants of 10µM, our values of $K_{r,b}$ imply an effective local concentration of 0.33mM. Smaller values of $K_{r,b}$ would lead to a stronger dependence of receptor activity on the ratio of localized enzymes and correspondingly higher predictions for the level of signaling noise.

For given values of the catalytic rates, $K_r$ and $K_b$ set the values of the binding rates of tethered enzymes to the modification sites and the rates of unbinding from the modification sites. Additionally, the values of $K_{r,b}$ constrain the values of the tether unbinding rates $d_{r,b}^t$, as discussed in the main text. Suitable choices of these rates ensure that the numbers of cytoplasmic and localized enzymes are comparable [5] and that the steady-state receptor activity is somewhat robust to variations in the expression levels of CheR and CheB. Parameters related to CheB phosphorylation were primarily based on measured values (see below).

Given values of the above parameters, the catalytic rates $k_r$ and $k_b$ of methylation and demethylation were calibrated by comparison with the measured responses of cell populations to exponential time-varying ramps of chemoattractant (Figs. 2A and S2) (Ref. [4], 32°C data). Using a CheY-CheZ FRET pair, these measurements quantified changes in receptor activity in response to ramps of methyl-aspartate. During stimulus, activity tended to reach steady-state values dependent on the speed of the ramp. These steady-state values were determined by fitting the time trace of activity to an exponential decay (Fig. S2). Our model agrees well with the experimental results over a wide range of activity (Fig. 2A), but diverges for strong negative ramps because it does not include nonlinear CheB phosphorylation (see below).

The remaining binding rates of Table S2 were chosen to be consistent with those discussed above by requiring $a_r^t/a_r^m = a_{r*}^t/a_{r*}^m$, $a_b^t/a_b^m = a_{b*}^t/a_{b*}^m$, $a_r^t/a_b^t = a_{r*}^t/a_{b*}^t$, $a_r^m/a_b^m = a_{r*}^m/a_{b*}^m$, which satisfies detailed balance.

### 3. Parameter values for models without enzyme localization

Parameter values in Table S4 are used in the analytical model without enzyme localization and are taken from a previous study, in which they were calibrated to fit experimental measurements [4]. Parameter values (Table S5) for the numerical implementation of this model (**B1**) were derived from these values.

| | | |
|---|---|---|
| Total number of receptor monomers (base) | $T_{Tot,0}$ | 14400 |
| Total number of CheR (base) | $R_{Tot,0}$ | 140 |
| Total number of CheB (base) | $B_{Tot,0}$ | 240 |
| Number of receptor dimers per MWC cluster | $N$ | 6 |
| Basal free energy difference, active and inactive cluster | $\varepsilon_0$ | 6 (units of $k_BT$) |
| Free energy change per added methyl group | $\varepsilon_1$ | 1 (units of $k_BT$) |

| MeAsp dissociation constant, inactive Tar receptor | $K$ | 0.0182 mM |
|---|---|---|
| MeAsp dissociation constant, active Tar receptor | $K^*$ | 3 mM |

**Table S1**: Parameter names and values common to all models.

| Bulk CheR binding to tether | $a_r^t$ | 1/14.7 s$^{-1}$/$T_{Tot,0}$ |
|---|---|---|
| Bulk CheR binding to modification site | $a_r^m$ | 0.0245 s$^{-1}$/$T_{Tot,0}$ |
| Tethered CheR binding modification site, same receptor | $a_{r*}^m$ | 400 s$^{-1}$ |
| CheR binding tether while attached to mod. site, same receptor | $a_{r*}^t$ | 1110 s$^{-1}$ |
| Tethered CheR binding modification site, neighboring receptor | $a_{r*}^{m'}$ | 400 s$^{-1}$ |
| CheR binding tether while attached to mod. site, neighboring rec. | $a_{r*}^{t'}$ | 1110 s$^{-1}$ |
| CheR unbinding modification site | $d_r^m$ | 9.3 s$^{-1}$ |
| CheR unbinding tether site | $d_r^t$ | 5 s$^{-1}$ |
| CheR catalytic rate | $k_r$ | 2.7 s$^{-1}$ |
| Bulk CheB-P binding to tether | $a_b^t$ | 1/16.3 s$^{-1}$/$T_{Tot,0}$ |
| Bulk CheB-P binding to modification site | $a_b^m$ | 0.0245 s$^{-1}$/$T_{Tot,0}$ |
| Tethered CheB-P binding modification site, same receptor | $a_{b*}^m$ | 400 s$^{-1}$ |
| CheB-P binding tether while attached to mod. site | $a_{b*}^t$ | 1000 s$^{-1}$ |
| Tethered CheB-P binding modification site, neighboring receptor | $a_{b*}^{m'}$ | 400 s$^{-1}$ |
| CheB-P binding tether while attached to mod. site, neighboring rec. | $a_{b*}^{t'}$ | 1000 s$^{-1}$ |
| CheB-P unbinding modification site | $d_b^m$ | 9 s$^{-1}$ |
| CheB-P unbinding tether site | $d_b^t$ | 5 s$^{-1}$ |
| CheB-P catalytic rate | $k_b$ | 3 s$^{-1}$ |
| CheB phosphorylation rate | $a_p$ | 3 s$^{-1}$/ $T_{Tot}$ |
| CheB-P dephosphorylation rate | $d_p$ | 0.37 s$^{-1}$ |

**Table S2**: Parameter values for stochastic simulation of model **M1** with enzyme localization. Rates are designated as in Fig. 1B with an *r* or *b* subscript to denote rates of CheR and CheB reactions.

| CheR catalytic rate | $k_r$ | |
|---|---|---|
| CheB catalytic rate | $k_b$ | $2N (0.03\ T_{Tot}/R_{Tot})$ s$^{-1}$ |
| CheR unbinding modification site | $d_r$ | |
| CheB unbinding modification site | $d_b$ | |
| CheR binding to modification site | $a_r$ | $(k_r + d_r)/(0.43\ T_{Tot})$ |
| CheB binding to modification site | $a_b$ | $(k_b + d_b)/(0.3\ T_{Tot})$ |

**Table S3:** Parameter values for stochastic simulation of the model **B1** with no enzyme localization.

| Bulk CheR binding to tether | $a_r^t$ | 1/14.7 s⁻¹/$T_{Tot,0}$ |
|---|---|---|
| CheR unbinding tether | $d_r^t$ | 5 s⁻¹ |
| Bulk CheB binding to tether | $a_b^t$ | 1/16.3 s⁻¹/$T_{Tot,0}$ |
| CheB unbinding tether | $d_b^t$ | 5 s⁻¹ |
| CheR catalytic rate | $k_r$ | 2.7 s⁻¹ |
| CheB catalytic rate | $k_b$ | 3 s⁻¹ |
| Tethered CheR modification site affinity | $K_r$ | 0.03 |
| Tethered CheB modification site affinity | $K_b$ | 0.03 |
| CheB phosphorylation rate | $a_p$ | 3 s⁻¹/ $T_{Tot}$ |
| CheB-P dephosphorylation rate | $d_p$ | 0.37 s⁻¹ |

**Table S4**: Parameter values for mean-field analytical model with enzyme localization. All values are derived from values of corresponding parameters in the numerical model **M1** (Table S2).

| CheR catalytic rate | $k_r$ | 2N (0.03 $T_{Tot}/R_{Tot}$) s⁻¹ |
|---|---|---|
| CheB catalytic rate | $k_b$ | 2N (0.03 $T_{Tot}/B_{Tot}$) s⁻¹ |
| Tethered CheR modification site affinity | $K_r$ | 0.43 $T_{Tot}$ |
| Tethered CheB modification site affinity | $K_b$ | 0.3 $T_{Tot}$ |

**Table S5**: Parameter values for analytical version of model **B1** with no enzyme localization.

| Model | Base model | Parameters changed from base model |
|---|---|---|
| **M2** | **M1** | $a_{r*}^{m'} = a_{r*}^{t'} = a_{b*}^{m'} = a_{b*}^{m'} = 0$ |
| **M3** | **M1** | $d_r^t = d_b^t = 0.25$ s⁻¹ |
| **B2** | **B1** | $a_r$, $a_b$ increased tenfold |
| **B2** (analytical) | **B1** (analytical) | $K_r$, $K_b$ increased tenfold |

**Table S6:** Changes in parameter values for the derived models **M2**, **M3**, and **B2**.

**CheB phosphorylation**

CheB is phosphorylated by the kinase CheA. Since only phosphorylated CheB is able to efficiently dock with and demethylate chemoreceptors [5], this arrangement constitutes a negative feedback loop. We implemented a simple CheB phosphorylation loop in our numerical models (**M1**, **M2**, **M3**) and in the analytical model with enzyme localization, following previous theoretical studies [8-11]. We model CheB phosphorylation through the reaction T + CheB → T + CheB-P with rate $a_p \cdot a(T)$ in which $a(T)$ is the activity of the receptor T. The rate of CheB-P autodephosphorylation $d_p$ has been measured by previous independent studies with excellent agreement [12,13]. We estimated the maximum phosphorylation rate $a_p$ using a simple model that considered CheA, CheB and CheY phosphorylation. Let active CheA autophosphorylate with rate $k_a$, CheA-P ($A_p$) phosphorylate CheY with rate $k_y$ = 100 $\mu M^{-1}$ $s^{-1}$ [14], and CheY-P ($Y_p$) dephosphorylate via CheZ with rate $k_z$ = 3 $s^{-1}$ [12]. Then at steady state

$$ak_a\left(A_{Tot} - A_p\right) \sim k_y A_p \left(Y_{Tot} - Y_p\right)$$

$$k_y A_p \left(Y_{Tot} - Y_p\right) = k_z Y_p,$$

in which $A_{Tot}$ = 5.3 $\mu M$ and $Y_{Tot}$ = 9.7 $\mu M$ are the total concentrations of CheA and CheY [3] for a cell volume of 1.4 fL [15]. Solving these equations with the requirement that $Y_p$ = 2.6 $\mu M$ at $a$ = 0.5 implies that $k_a \sim 3$ $s^{-1}$, which in turn implies that $A_p \sim 0.2$ $\mu M \times a$ for most values of $a$. Since CheA-P phosphorylates CheB with rate 15 $\mu M^{-1}$ $s^{-1}$ [14], this estimate implies that $a_p$ = 3 $s^{-1}$ /$T_{Tot}$. In our implementation, then, the fast phosphorylation of CheA is effectively at steady state, an approximation that serves to reduce the number of parameters and significantly speed simulation.

*In vivo* measurements of activity (as measured through the CheY-CheZ interaction via FRET) suggest that the rate of CheB phosphorylation has nonlinear dependence on kinase activity. Specifically, this dependence is inferred from: (1) significant asymmetry in the rate of the adaptation response to positive and negative step stimuli [16]; (2) a sharp increase in the demethylation rate at high kinase activities, measured through stimulation by exponential time-varying ramps of chemoattractant [4]. Comparison of results from the ramp stimulus experiments to a theoretical model indicates that CheB phosphorylation only affects the adaptation kinetics at high activities ($a > 0.74$). Since the molecular processes underlying this nonlinearity are currently unknown, we included simpler, linear CheB feedback in our models. As a result, while our model agrees well with measurements over a wide range of activities, it deviates at the highest activities (Fig. 2A). These activities are higher than the mean activities of typical, unstimulated wild-type cells, which are the primary focus of our study.

**Analytical models**

*1. Fluctuations without enzyme localization*

The model in this section is an analytical treatment of **B1**. We consider a model of MWC signaling complexes each consisting of $N$ = 6 receptor dimers. Let $m$ be the average methylation level per signaling complex, varying between 0 and $8N$. We begin by studying the dynamics of the total methylation level $M = mT_{Tot}/2N$, the total number of methyl groups bound to receptors. To the equation of motion for the mean of $M$, we can apply the linear noise approximation (LNA) directly to obtain a stochastic differential equation for the dynamics of $M$. From this basis, we can then calculate fluctuations in both $m$ and the fraction of active receptor clusters $a$.

In this model we neglect the localization of the cytoplasmic enzymes with the tether site and simply assume that CheR binds the receptor modification site with affinity $(1-a)/K_r$ and CheB binds with affinity $a/K_b$. Then the number of CheR-receptor complexes is $RT/K_r$ and CheB-receptor complexes is $BT^*/K_b$, in which $R$ and $B$ are the numbers of free enzymes and $T$ and $T^*$ are the numbers of inactive and active receptors. $K_r$ and $K_b$ are Michaelis-Menten constants. Then for catalytic rates $k_r$ and $k_b$, the change in $M$ is given by $dM/dt = k_r RT/K_r - k_b BT^*/K_b$. Using the conservation of enzymes $R_{Tot} = R(1 + T/K_r)$ and $B_{Tot} = B(1 + T^*/K_b)$ yields the equation

$$\frac{dM}{dt} = \frac{k_r R_{Tot} T}{K_r + T} - \frac{k_b B_{Tot} T^*}{K_b + T^*}, \quad (S2)$$

which has appreared in numerous previous models.

Using the LNA, we convert Eq. (S2) into the stochastic differential equation $dM = \dot{M} dt + \sqrt{D_M}\, dW$, in which $W(t)$ is the Wiener process. The relevant noise intensity is

$$D_M = \frac{k_r R_{Tot} T}{K_r + T} + \frac{k_b B_{Tot} T^*}{K_b + T^*}. \quad (S3)$$

Since we are ultimately considering small fluctuations around the steady state, the quantities appearing in $D_M$ are evaluated at their steady-state values.

To calculate the variation in the fraction of active receptors

$$a(m) = T^*(1 + B/K_b)/T_{Tot}, \quad (S4)$$

we apply Itō's Lemma to the function $a(m)$, yielding

$$da = \frac{\partial a}{\partial m}\frac{dM}{T_{Tot}/2N} + \frac{1}{2}\frac{\partial^2 a}{\partial m^2}\frac{D_M}{(T_{Tot}/2N)^2}dt = \frac{2N}{T_{Tot}}\frac{\partial a}{\partial m}\left[\frac{k_r R_{Tot} T}{K_r + T}dt - \frac{k_b B_{Tot} T^*}{K_b + T^*}dt + \sqrt{D_M}\, dW\right]$$
$$+ \frac{1}{2}\frac{\partial^2 a}{\partial m^2}\frac{D_M}{(T_{Tot}/2N)^2}dt.$$

(S5)

The derivatives of receptor activity $a(m, L)$, Eq. (13) of the main text, with respect to the methylation level per cluster $m$ may be written in terms of the activity as

$$\frac{\partial a}{\partial m} = \varepsilon_1 a(1-a) > 0$$

$$\frac{\partial^2 a}{\partial m^2} = \varepsilon_1^2 a(1-a)(1-2a). \quad (S6)$$

Using Eqs. (S3-6) and the conservation of receptor number

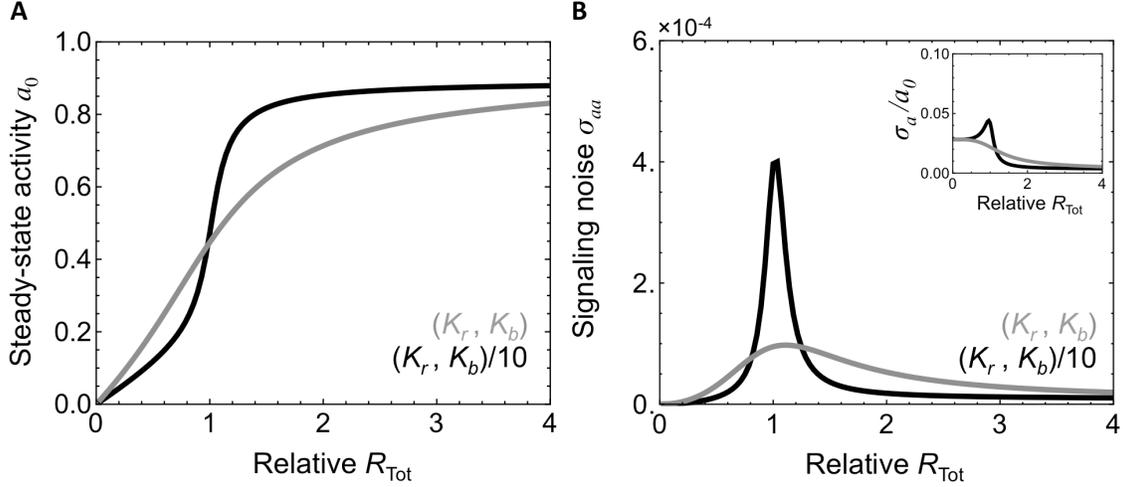

**Fig. S3**: Fluctuations in the analytical model with no enzyme localization. The noise level within a narrow range of CheR values increases as the dependence of the steady-state activity on CheR count becomes steeper. (A) Steady state activity $a_0$ as a function of normalized CheR count for the parameters used in Fig. 4 (gray) and with Michaelis-Menten constants $K_r$ and $K_b$ reduced by a factor of 10 (black). The latter curve exhibits an extreme dependence on variations in CheR count. (B) Variance $\sigma_{aa}$ and relative noise $\sigma_a/a_0$ (inset) in activity at the steady state as a function of normalized CheR count for original (gray) and reduced $K_r$ and $K_b$ (black). Reducing $K_r$ and $K_b$ increases the relative noise level to nearly 5%.

$$T_{Tot} = T\left(1 + \frac{R_{Tot}}{K_r + T}\right) + T^*\left(1 + \frac{B_{Tot}}{K_b + T^*}\right), \qquad (S7)$$

we can express $da$ in terms of only $a$ and constants. Linearizing the resulting expression around the steady state of $a$ yields an equation of the form $d(\delta a) = -(\delta a/\tau)dt + \sqrt{D_a}\,dW$ from the which the expression for the relaxation time can be read off. The variance of this process is $\mathrm{var}(\delta a) = \sigma_{aa} = \tau D_a/2$ with $D_a = (\partial a/\partial m)^2 D_M/(T_{Tot}/2N)^2$. For receptor activity $a < 0.74$ and the parameters in Table S4, this model is equivalent to a recent model calibrated from population responses to exponential ramps of attractant [4]. For $a > 0.74$, that model included a nonlinear CheB-P feedback term, which we have neglected here for simplicity and since we are primarily interested in comparisons at lower activities where signaling noise is higher.

As previously discussed [8], the noise level $\sigma_{aa}$ increases with the dependence of the activity at steady state on the number of CheR, $a_0(R_{Tot})$. In Fig. S3A, $a_0(R_{Tot})$ is plotted both for the Michaelis-Menten constants in Table S3 and reduced by a factor of 10 (Table S6). As known from Goldbeter and Koshland [17], reducing $K_r$ and $K_b$ steepens $a_0(R_{Tot})$ and correspondingly increases the noise level in activity (Fig. S3B). This model with increased affinities corresponds to the numerical model **B2**.

## 2. Fluctuations in the enzyme localization model

The dynamics of the analytical model with enzyme localization are given by the stochastic Eqs. (S8-10), adapted here from the chemical Langevin Eqs. (8-10) of the main text

$$dM = \left[ \frac{k_r R^*_{Tot}(1-a)}{K_r + 1 - a} - \frac{k_b B^*_{p,Tot} a}{K_b + a} \right] dt + \sqrt{D_M} dW_M \quad (S8)$$

$$dR^*_{Tot} = \left[ a^t_r T_{Tot} R - d^t_r \frac{K_r}{K_r + 1 - a} R^*_{Tot} \right] dt + \sqrt{D_r} dW_r \quad (S9)$$

$$dB^*_{p,Tot} = \left[ a^t_b T_{Tot} B_p - d^t_b \frac{K_b}{K_b + a} B^*_{p,Tot} \right] dt + \sqrt{D_b} dW_b, \quad (S10)$$

in which $W_i(t)$ are independent Wiener processes. The derivation is given in the main text. Additionally, enzyme numbers are conserved according to $R_{Tot} = R + R^*_{Tot}$ and $B_{Tot} = B + B_p + B^*_{p,Tot}$. Applying Itō's lemma to Eq. (S8), we find that receptor activity evolves according to

$$da = \frac{2N}{T_{Tot}} \frac{\partial a}{\partial m} \left[ \frac{k_r R^*_{Tot}(1-a)}{K_r + 1 - a} dt - \frac{k_b B^*_{p,Tot} a}{K_b + a} dt + \sqrt{D_M} dW \right] + \frac{1}{2} \frac{\partial^2 a}{\partial m^2} \frac{D_M}{(T_{Tot}/2N)^2} dt. \quad (S11)$$

The derivatives of activity $a$ with respect to $m$ may be expressed using Eq. (S6) as in the previous section. Equations (S9-11) may then be rewritten in the form

$$d\mathbf{X} = \mathbf{F}(\mathbf{X}) dt + B d\mathbf{W}, \quad (S12)$$

in which $\mathbf{X} = (a, R^*_{Tot}, B^*_{p,Tot})^T$ and the components of the diffusion matrix $B$ are calculated by the LNA [18,19]

$$B = \text{diag}\left( \frac{2N}{T_{Tot}} \frac{\partial a}{\partial m} \left[ \frac{k_r R^*_{Tot}(1-a)}{K_r + 1 - a} + \frac{k_b B^*_{p,Tot} a}{K_b + a} \right]^{1/2}, \left[ a^t_r T_{Tot} R + d^t_r \frac{K_r}{K_r + 1 - a} R^*_{Tot} \right]^{1/2}, \left[ a^t_b T_{Tot} B_p + d^t_b \frac{K_b}{K_b + a} B^*_{p,Tot} \right]^{1/2} \right)$$

(S13)

and evaluated at the steady state, $\mathbf{F}(\mathbf{X}_0) = 0$. To consider small deviations $\mathbf{x} = \mathbf{X} - \mathbf{X}_0$, we linearize Eq. (S12) by calculating the Jacobian $A$ of $\mathbf{F}$. The resulting linear system

$$d\mathbf{x} = A\mathbf{x} dt + B d\mathbf{W} \quad (S14)$$

is a multivariate Ornstein-Uhlenbeck process. The steady state variance in the output of the system is obtained by solving the Lyapunov equation [20,21]

$$A\sigma + \sigma A^T + B^T B = 0 \quad \text{(S15)}$$

for the covariance matrix $\sigma$. The autocorrelation matrix $C$ at steady state may also be calculated by

$$C(t) = \exp(At)\sigma. \quad \text{(S16)}$$

*3. Detailed model of enzyme localization*

The analytical model of the previous section incorporates the features responsible for increased signaling noise when enzyme localization and brachiation are included in our simulation of the bacterial chemotaxis system. However, it also overestimates the magnitude of this noise. Two reasons exist for this overestimate. First, the model assumes that all receptors are equally accessible to all localized enzymes, meaning that methylation is fully distributive. In this picture, enzymes bind the receptors for which they have the highest affinity (low methylation level for CheR and high methylation level for CheB) regardless of the state of the receptor to which they are tethered. The result is an overestimate of the binding affinity of localized enzymes for the receptor substrate. Second, the previous analytical model does not consider a distribution of methylation levels. Rather, $\partial a / \partial m$ is evaluated at only a single methylation level corresponding to the mean activity of the system. This approximation tends to overestimate $\partial a / \partial m$, most significantly for systems with mean activity of $a \sim 0.5$, where $\partial a / \partial m$ is largest. Addressing these issues requires a model that considers the dynamics of MWC complexes of each methylation level $m$ individually. Additionally, the model must track the numbers of enzymes localized to complexes at each methylation level. To tune the processivity of methylation, we introduce a parameter $\beta$ representing the rate at which localized enzymes randomize their position. The value $\beta = 0$ corresponds to completely processive methylation and the limit $\beta \to \infty$ corresponds to purely distributive methylation, reducing the model to Eqs. (8-10) of the main text.

Let $T_{Tot, m}$ be the total number of receptor monomers within MWC signaling complexes of methylation level $m$. Also let $R_m^*$ and $B_{p,m}^*$ be the number of CheR and CheB-P localized within clusters of methylation level $m$ but not bound to modification sites. For simplicity, we consider only the case of a tethered enzyme binding a modification site in the same cluster that it is localized. This binding of the modification site forms complexes denoted by $\overline{R_m^* T_m}$ and $\overline{B_{p,m}^* T_m}$. The changes in the number of these complexes due to modification site (un)binding and catalysis is then:

$$\frac{d}{dt} \overline{R_m^* T_m} = a_{r*}^m (1 - a_m) R_m^* - (d_r^m + k_r) \overline{R_m^* T_m}$$

$$\frac{d}{dt} B^*_{p,m} T_m = a^m_{b*} a_m B^*_{p,m} - \left(d^m_b + k_b\right) \overline{B^*_{p,m} T_m} \qquad \text{(S17)}$$

Taking the approximation $dR^*_m T_m/dt = dB^*_{p,m} T_m/dt = 0$ as in Eqs. (5-7) of the main text, changes in $T_{Tot,m}$ are given by

$$\frac{d}{dt} T_{Tot,0} = -\frac{k_r}{K_r} R^*_0 (1-a_0) + \frac{k_b}{K_b} B^*_{p,1} a_1$$

$$\vdots$$

$$\frac{d}{dt} T_{Tot,m} = \frac{k_r}{K_r} R^*_{m-1}(1-a_{m-1}) - \frac{k_r}{K_r} R^*_m (1-a_m) - \frac{k_b}{K_b} B^*_{p,m} a_m + \frac{k_b}{K_b} B^*_{p,m+1} a_{m+1} \qquad \text{(S18)}$$

$$\vdots$$

$$\frac{d}{dt} T_{Tot,8N} = \frac{k_r}{K_r} R^*_{8N-1}(1-a_{8N-1}) - \frac{k_b}{K_b} B^*_{p,8N} a_{8N}$$

in which the cluster activity $a_m = a(m, L)$. All parameters are defined in the same manner as those for the previous analytical model given in Table S4.

We now write the analogs of Eqs. (1, 2) in the main text, which here describe the binding and unbinding of cytoplasmic CheR ($R$) and CheB-P ($B_p$) to tether sites within signaling complexes of methylation level $m$. The quantities

$$R^*_{Tot,m} = R^*_m \left(1 + \frac{1-a_m}{K_r}\right),$$

$$B^*_{p,Tot,m} = B^*_{p,m} \left(1 + \frac{a_m}{K_b}\right) \qquad \text{(S19)}$$

denote the total numbers of localized enzymes at each methylation level. These total localized enzyme counts evolve according to

$$\frac{d}{dt} R^*_{Tot,m} = a^t_r R T_{Tot,m} - d^t_r R^*_m + \frac{k_r}{K_r}\left[R^*_{m-1}(1-a_{m-1}) - R^*_m(1-a_m)\right] + \beta\left(-R^*_m + R^* \frac{T_{Tot,m}}{T_{Tot}}\right)$$

(S20)

$$\frac{d}{dt} B^*_{p,Tot,m} = a^t_b B_p T_{Tot,m} - d^t_b B^*_{p,m} + \frac{k_b}{K_b}\left(-B^*_{p,m} a_m + B^*_{p,m+1} a_{m+1}\right) + \beta\left(-B^*_{p,m} + B^*_p \frac{T_{Tot,m}}{T_{Tot}}\right)$$

(S21).

The first term two terms of the right hand side arise from the binding and unbinding of localized enzymes from the tether site. As in the previous model, we here assume the number of free tethers to be large compared to the number of localized enzymes and therefore $\sim T_{Tot,m}$. The third and fourth terms (multiplied by $k_{r,b}/K_{r,b}$) represent the localized enzyme (de)methylating the complex within which it is localized. The last two terms introduce a redistribution of localized enzymes between the receptor clusters occurring at a rate $\beta$. The probability of a given complex receiving a new enzyme is proportional to its relative abundance $T_{Tot,m}/T_{Tot}$.

Molecule counts are conserved according to

$$R_{Tot} = R + \sum_m R^*_{Tot,m}$$

$$B_{Tot} = B + B_p + \sum_m B^*_{p,Tot,m} \qquad (S22)$$

$$T_{Tot} = \sum_m T_{Tot,m}$$

and the overall activity $a$ is calculated according to

$$a = \frac{1}{T_{Tot}} \sum_m a_m T_{Tot,m}. \qquad (S23)$$

In the limit $\beta \rightarrow \infty$, Eqs. (S18, S20, S21) reduce to Eqs. (8-10) of the main text. Summing Eqs. (S20, S21) over $m$ and using the definitions $R^* = \sum_m R^*_m$ and $B^*_p = \sum_m B^*_{p,m}$ reduces them to Eqs. (9, 10) describing the dynamics of the total number of localized enzymes. To recover Eq. (8), we note that as $\beta \rightarrow \infty$, it follows from Eqs. (S20, S21) that $R^*_m \sim R^* T_{Tot,m}/T_{Tot}$ and $B^*_{p,m} \sim B^*_p T_{Tot,m}/T_{Tot}$. Inserting these values into Eq. (S18) and summing $dm/dt \sim \sum_m m \dot{T}_{Tot,m}/T_{Tot}$ yields Eq. (8) for $a(m = 0) \sim 0$ and $a(m = 8N) \sim 1$ [8]. This limit corresponds to fully distributive methylation in which localized enzymes are completely redistributed between each methylation event according to the abundances $T_{Tot,m}$.

For calculations about the steady state with no stimulus, only methylation levels up to $m \sim 16$ must be considered, since $a(16, L = 0) \sim 1$ guarantees that CheR will not methylate clusters beyond this limit. Signaling noise in the overall activity $a$ is calculated by linearizing and applying the linear noise approximation to Eqs. (S18, S20, S21). The variance in the overall activity $\sigma_{aa}$ is calculated from the covariances of the $T_{Tot,m}$. Results of this calculation for $\beta = 0$ (fully processive) and $\beta = 20$ s$^{-1}$ (more distributive) are shown in Fig. S4. The more distributive model exhibits larger fluctuations and a higher affinity of the localized enzymes for the receptor substrate.

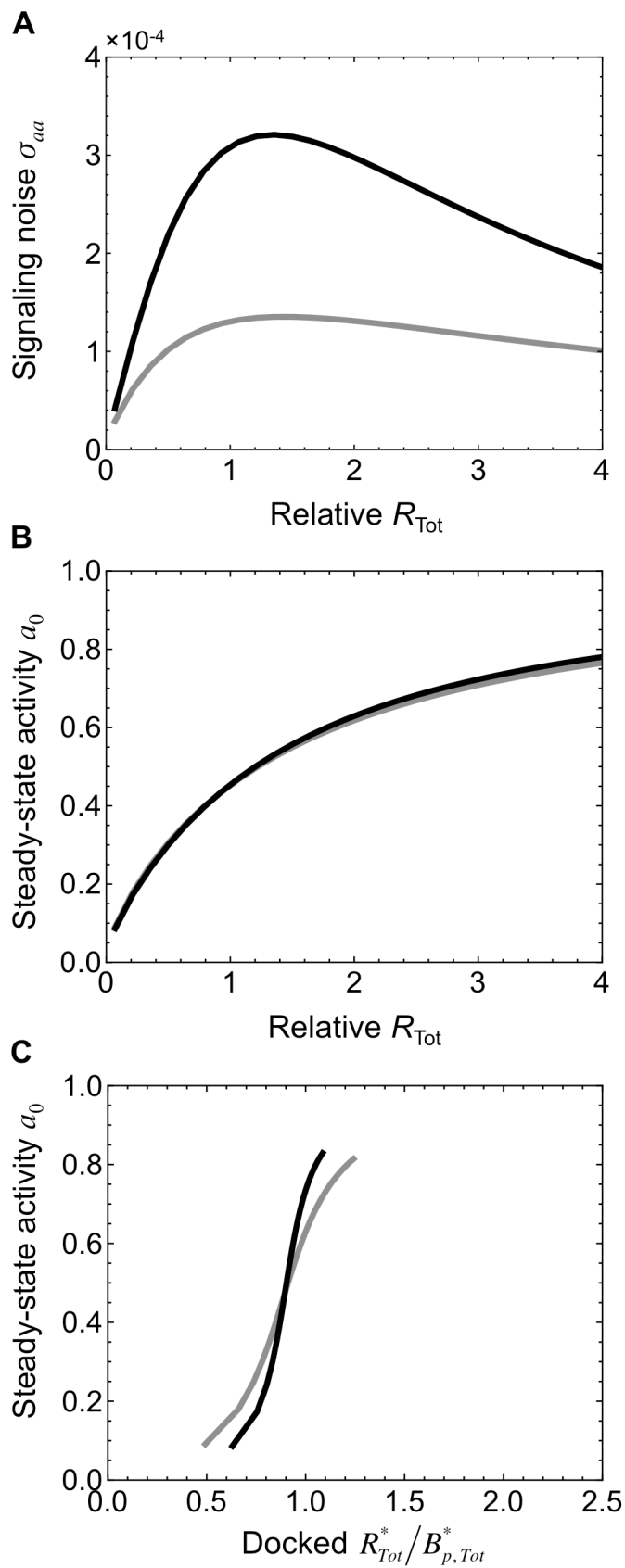

**Fig. S4**: Increasing the distributivity of methylation in the detailed analytical model (Text S1) increases noise and the affinity of localized enzymes for the receptor substrate. (A) Variance $\sigma_{aa}$ in overall activity as a function of total CheR count for fully processive methylation, $\beta = 0$ (gray), and more distributive methylation, $\beta = 20$ s$^{-1}$ (black) (B) The steady-state activity $a_0$ as a function of total CheR is similar for both $\beta = 0$ (gray) and $\beta = 20$ s$^{-1}$ (black). (C) Steady-state activity $a_0$ versus localized CheR/CheB-P, $R^*_{Tot}/B^*_{p,Tot}$, is much steeper in the more distributive model with $\beta = 20$ s$^{-1}$ (black) than $\beta = 0$ (gray).

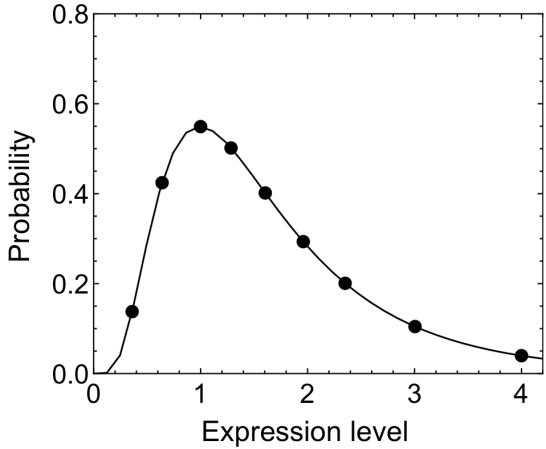

Fig. S5: Estimated distribution of overall chemotaxis protein expression levels in a wild-type population relative to the most common expression level. We sampled representative cells (points) from a population in which the ratio CheR/CheB/chemoreceptors is maintained while the overall expression level follows a log-normal distribution. Signaling noise levels for these representative cells are shown in Fig. 3A of the main text.

**Comparing the dependence of receptor activity on the localized enzyme ratio**

When calculating the localized CheR to CheB-P ratio for the numerical models (Fig. 4C, D), we ignore the population of "inert" CheR and CheB-P localized respectively within fully methylated or demethylated assistance neighborhoods or, for the model **M2** lacking assistance neighborhoods, tethered to fully methylated or demethylated receptor dimers. While these inert enzymes are able to bind the modification sites of the receptors, they are unable to participate in methylation-demethylation and therefore do not affect the activity of the receptor cluster. This consideration provides the best comparison to Fig. 4B, since the analytical model assumes all bound enzymes will (de)methylate receptors at the same rate, Eq. (7). The population of inert CheR is small for all models, but the fraction of inert CheB is high for the processive models since many receptors are fully demethylated (Fig. S6). We note that since the simulated MWC complexes in the absence of stimulus are half active when the methylation of the complex is $m = 6$ (out of a possible 48) and almost fully active with $m = 12$, fully demethylated receptors are common even when the activity of the receptor cluster is high.

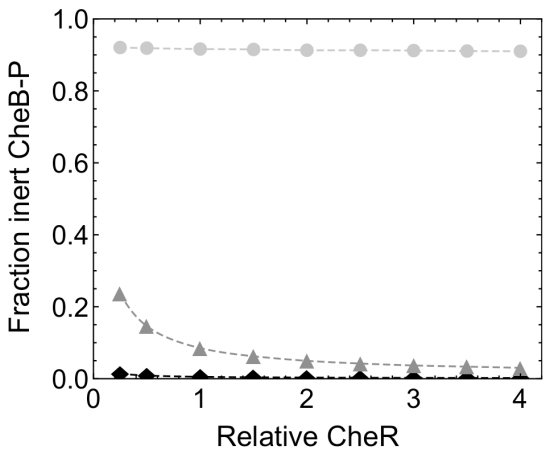

Fig. S6: Mean fraction of "inert" CheB-P tethered within fully demethylated assistance neighborhoods (for models **M1**, black, and **M3**, light gray) or with fully demethylated receptor dimers (**M2**, dark gray) versus total CheR. These enzymes may bind the modification sites of receptors but will be unable to demethylate once bound. These enzymes are unable to affect the activity of the receptor cluster and are therefore not counted when calculating the ratio of localized CheR to CheB-P for Fig. 4. Since very few receptors are fully methylated, the number of inert, localized CheR is negligible (< 1) for all models. This situation arises because MWC signaling complexes are highly active even at low methylation levels: in the absence of stimulus, $a = 0.5$ for $m = 6$ (out of 48) and $a \sim 1$ for $m = 14$. Consequently, many receptor dimers are fully demethylated even for cases in which the average receptor activity is high. In contrast, fully methylated dimers are rare.

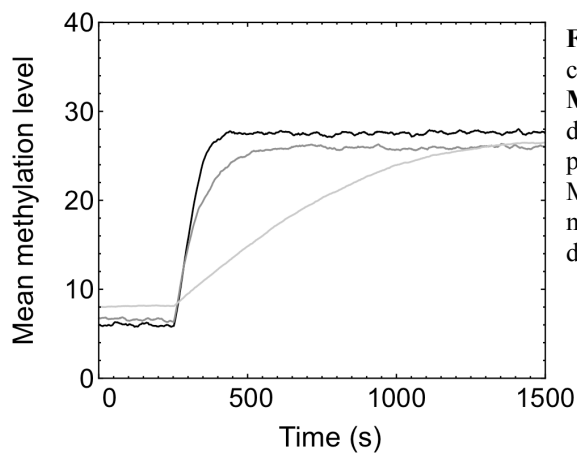

**Fig. S7**: Average methylation level per MWC complex as a function of time for numerical models **M1** (black), **M2** (light gray), and **M3** (dark gray) during the simulations shown in Fig. 2B (lower panel) of the main text. A step stimulus of 1 mM MeAsp was presented at 200 s. The most distributive model **M1** displays the highest methylation rate during the adaptation process.

**Comparing adaptation rates between the numerical models**

When presented with a large attractant stimulus, the numerical models clearly displayed different rates of adaptation with the most distributive model **M1** reaching its adapted level of activity first (Fig. 2B). Fig. S7 shows the mean methylation level per MWC complex versus time for the simulations plotted in Fig. 2B, which clearly shows that **M1** (black) has the highest overall methylation rate during adaptation. We note that since the enzymes in model **M3** have lower rates of tether unbinding, ~30% more enzymes are localized to the cluster than in **M1**. This leads to the slightly faster initial rate of methylation for **M3** compared to **M1**. This rate, however, drops due to the processivity of methylation in **M3**, as localized CheR is unable to escape from pockets of high methylation. In contrast, the methylation rate of the most distributive model **M1** remains nearly constant during adaptation.

| Model | **M1** | **M2** | **M3** |
|---|---|---|---|
| CheR sampling rate (dimers/s) | 4.1 | 1.1 | 0.64 |
| CheB-P sampling rate (dimers/s) | 3.4 | 0.26 | 0.36 |

**Table S7**: Number of unique dimers visited by localized enzymes per second for the numerical models. Higher rates indicate more distributive methylation.